\documentclass[letterpaper,11pt]{article}
\pdfoutput=1

\usepackage{jheppub}
\usepackage{subfig}

\def\Fig#1{fig.~{\ref{#1}}}
\def\eq#1{eq.~(\ref{#1})}
\def\eqs#1#2{eqs.~(\ref{#1}) and~(\ref{#2})}
\def\App#1{Appendix~\ref{#1}}
\DeclareRobustCommand{\Sec}[1]{sec.~\ref{#1}}
\DeclareRobustCommand{\Eq}[1]{eq.~(\ref{#1})}

\def\cE{\mathcal{E}}
\def\cN{\mathcal{N}}

\newcommand{\al}{\alpha}
\newcommand{\ga}{\gamma}
\newcommand{\Ga}{\Gamma}
\newcommand{\de}{\delta}
\newcommand{\De}{\Delta}
\newcommand{\eps}{\epsilon}
\newcommand{\si}{\sigma}

\def \as {\relax\ifmmode\alpha_s\else{$\alpha_s${ }}\fi}

\def\img{{\rm i}}
\newcommand{\df}{\mathrm{d}}
\newcommand{\nn}{\nonumber}

\allowdisplaybreaks[2]

\begin{document}

\title{Renormalization Group Flows for Track Function Moments}

\author[1,2]{Max Jaarsma,}
\author[3]{Yibei Li,}
\author[4]{Ian Moult,}
\author[1,2]{Wouter Waalewijn,}
\author[3]{and Hua Xing Zhu}
\affiliation[1]{Nikhef, Theory Group,
	Science Park 105, 1098 XG, Amsterdam, The Netherlands}
\affiliation[2]{Institute for Theoretical Physics Amsterdam and Delta Institute for 
 Theoretical Physics, University of Amsterdam, Science Park 904, 1098 XH Amsterdam, The Netherlands}
\affiliation[3]{Zhejiang Institute of Modern Physics, Department of Physics, Zhejiang University, Hangzhou, Zhejiang 310027, China}
\affiliation[4]{Department of Physics, Yale University, New Haven, CT 06511, USA\vspace{0.5ex}}

\abstract{Track functions describe the collective effect of the fragmentation of quarks and gluons into charged hadrons, making them a key ingredient for jet substructure measurements at hadron colliders, where track-based measurements offer superior angular resolution. The first moment of the track function, describing the average energy deposited in charged particles, is a simple and well-studied object. However, measurements of higher-point correlations of energy flow necessitate a characterization of fluctuations in the hadronization process, described theoretically by higher moments of the track function. 
In this paper we derive the structure of the renormalization group (RG) evolution equations for track function moments. We show that energy conservation gives rise to a shift symmetry that allows the evolution equations to be written in terms of cumulants, $\kappa(N)$, and the difference between the first moment of quark and gluon track functions, $\Delta$. The uniqueness of the first three cumulants then fixes their all-order evolution to be DGLAP, up to corrections involving powers of $\Delta$, that are numerically suppressed by an effective order in the perturbative expansion for phenomenological track functions. However, at the fourth cumulant and beyond there is non-trivial RG mixing into products of cumulants such as $\kappa(4)$ into $\kappa(2)^2$.
We analytically compute the evolution equations up to the sixth moment at $\mathcal{O}(\alpha_s^2)$, and study the associated RG flows. These results allow for the study of up to six-point correlations in energy flow using tracks, paving the way for precision jet substructure at the LHC.
}

\maketitle

\section{Introduction}

The characterization of energy flow within jets, colloquially known as jet substructure, provides new ways to study QCD and search for potential new physics at the LHC \cite{Larkoski:2017jix,Marzani:2019hun}. The remarkable advances in this area in the last decade have primarily focused on the calculation of infrared and collinear (IRC) safe observables that can be computed within perturbative QCD, up to power corrections. The famous theorems of Kinoshita, Lee and Nauenberg \cite{Kinoshita:1962ur,Lee:1964is} state that this is only possible if one is completely inclusive over hadron species. As a consequence, such calculations can only be used to describe observables constructed from energy flow information, disregarding all the interesting information contained in other particle properties. Theoretically, these observables are therefore (combinations of) correlation functions of energy flow operators, $\langle \mathcal{E}(\vec n_1) \mathcal{E}(\vec n_2) \cdots \mathcal{E}(\vec n_k) \rangle$.

There is significant motivation to go beyond this energy flow paradigm, both for allowing more detailed tests of QCD, and for sharpening our tools in new physics searches. Such observables are inherently non-perturbative, as they require knowledge of the spectrum of hadrons in the theory. For example, at the LHC, many precision jet substructure measurements are made using tracks (charged particles), due to the improved angular resolution of the tracking system.  This sensitivity to hadronization can of course also be viewed as a positive if the goal is to understand features of the hadronization process. For example, the study of energy flow on charged or strange particles provides insight into how these quantum numbers evolve in the confinement process. 

The departure from IRC safety should not be done arbitrarily, and in particular, one should attempt to maintain the wealth of theoretical structures and advances of perturbative quantum field theory, but generalize them to a wider class of observables. In ref.~\cite{Li:2021zcf}, building on \cite{Chen:2020vvp}, it was shown that the natural way to extend the space of IRC safe observables to incorporate particle species information is to consider correlations of energy flow on subsets of particles. These are defined theoretically by considering an energy flow operator on a subset $R$ of particles, $\mathcal{E}_R(\vec n_1)$, and enable a much more general class of correlations to be studied, $\langle \mathcal{E}_{R_1}(\vec n_1) \mathcal{E}_{R_2}(\vec n_2) \cdots \mathcal{E}_{R_k}(\vec n_k) \rangle$, where in general the subsets, $R_i$, are distinct. As we will discuss, these observables exhibit a clean factorization into a non-perturbative component, and a perturbative component. The perturbative component shares many of the features of the standard energy correlators, and in particular can be computed at high perturbative orders using well-developed techniques from perturbative quantum field theory.  

Although the correlators $\langle \mathcal{E}_{R_1}(\vec n_1) \mathcal{E}_{R_2}(\vec n_2) \cdots \mathcal{E}_{R_k}(\vec n_k) \rangle$ cannot be directly computed in perturbation theory, they can be matched onto the standard energy flow correlators using non-perturbative track functions  \cite{Chang:2013rca,Chang:2013iba}. These track functions were introduced to describe the fraction of energy deposited into charged hadrons from a perturbative quark or gluon, however, they can trivially be generalized to the study of any other quantum number. Unlike standard fragmentation functions, track functions incorporate correlations between particles, arising from the fact that quarks and gluons can fragment into an arbitrary number of charged hadrons. As such, their evolution with scale is substantially more complicated, since all the correlations mix under evolution.

In ref.~\cite{Chen:2020vvp}, it was shown that, by restricting to correlation functions of energy flow measured on tracks, one is only sensitive to low moments of the track functions. These characterize the fluctuations in the hadronization process.\footnote{In analogy with the study of a spin system in statistical mechanics, the track function can be though of as the partition function or generating function, and its moments as the study of the expectations $\langle m^N \rangle$. Instead of studying the full renormalization group structure of the partition function, we consider the renormalization group of the low fluctuations, as is more standard.} To describe $N$-th order fluctuations requires only a finite set of operators, which mix under renormalization. Furthermore, the full track function distributions seem well-described by a truncated Gaussian, whose form is fixed by the first two moments.  In ref.~\cite{Li:2021zcf} it was shown that energy conservation places severe constraints on the RG evolution of the fluctuations, fixing the evolution of the first three moments to be DGLAP, up to corrections proportional to powers of $\Delta=T_q(1)-T_g(1)$. For track functions describing the production of electrically charged hadrons in QCD, $\Delta \ll 1$, effectively suppressing these contributions by an order in the perturbative expansion. At the fourth moment and beyond the fluctuations in the hadronization process exhibit non-trivial RG flows describing the mixing between different cumulants, for example $\kappa(4)$ and $\kappa(2)^2$. 

In this paper we discuss in detail the structure of the RG for the moments of the track functions.  In dimensional regularization, the corrections for the track functions are scaleless thus linking the evolution (UV poles) and the IR poles needed for incorporating track functions in calculations. We derive general constraints on the structure of the evolution that hold to all orders in perturbation theory, and in generic theories. In QCD, we then analytically compute the first six moments at next-to-leading order (NLO), and study the structure of their RG flows, which exhibit interesting mixing. For the first three moments the mixing terms are all suppressed by powers of $\Delta$ and smaller than the NNLO corrections, allowing us to extend our calculation to this order. We also argue, that due to the nonlinear nature of the track function evolution, it exhibits a UV fixed-point where the track functions become a delta function. Our explicit results enable the calculation of jet substructure observables sensitive to up to six point correlations in energy flow on tracks. 

While the primary motivation for this work is practical, namely enabling higher point correlators to be precisely measured at the LHC, the study of track functions is also of more formal theoretical interest. Track functions, and related multi-hadron fragmentation, are intrinsically Lorentzian observables whose RG evolution goes beyond standard DGLAP evolution. Although there has been significant recent progress in understanding certain classes of Lorentzian operators using lightray operators \cite{Kravchuk:2018htv}, this has primarily been restricted to operators on the leading Regge trajectory (which includes DGLAP). Understanding how the more general class of track function observables fits into this picture is interesting, and could lead to a better understanding of the analytic structure of Lorentzian observables in conformal field theories (CFTs). While we will not address this issue directly in this paper, our perturbative calculations provide important theoretical data for future investigations.

The outline of this paper is as follows: We discuss the flow of energy on subsets of particles in \Sec{sec:EEC}, motivating the study of moments of track functions. In \Sec{sec:GFF} we review the field-theoretic definition of track functions, and derive all-orders constraints on the renormalization group evolution of their moments.  We then restrict to NLO, and derive the specific constraints both for a pure gluon theory, as well as for QCD.  In \Sec{sec:results} we present results for the first six moments of the track functions at NLO, and describe the techniques used in the calculation. More details of the calculation for Pure Yang-Mills are given in app.~\ref{app:yang_mills}, which include results up to ninth moment, and the time-like splitting functions entering our results are collected in app.~\ref{app:splitting_moments}. In \Sec{sec:num} we numerically study the structure of the RG flows. We first show that in QCD, $\Delta \ll 1$, allowing us to extend our results for the evolution of the first three moments to NNLO. We then study the importance of non-linearities in the evolution of the fourth and fifth moments. We conclude in \Sec{sec:conc}.

\section{Energy Flow on Tracks and Track Function Moments}\label{sec:EEC}

To motivate the study of track function moments, we begin by reviewing the natural generalization of the study of correlations of energy flow, to the study of energy flow on subsets $R$ of particles. Here we will see that the non-perturbative information  required for this extension is precisely the moments of track functions, motivating our focus on these moments.

Energy flow in final states is characterized by the energy flow operator \cite{Sveshnikov:1995vi,Tkachov:1995kk,Korchemsky:1999kt,Bauer:2008dt,Hofman:2008ar,Belitsky:2013xxa,Belitsky:2013bja,Kravchuk:2018htv}
\begin{align}\label{eq:ANEC_op}
\mathcal{E}(\vec n) = \lim_{r\to \infty}  \int\limits_0^\infty \df t~ r^2 n^i T_{0i}(t,r \vec n)\,.
\end{align}
The canonical observables of the theory  are the $k$-point correlation functions  $\langle \mathcal{E}(\vec n_1) \mathcal{E}(\vec n_2) \cdots \mathcal{E}(\vec n_k) \rangle$. These generalize the original two-point correlator introduced early on in the QCD literature~\cite{Basham:1978bw}.  There has recently been significant interest in better understanding these observables from a number of different perspectives: These include higher loop perturbative calculations \cite{Belitsky:2013ofa,Dixon:2018qgp,Luo:2019nig,Henn:2019gkr}, resummation and effective field theory studies \cite{Moult:2018jzp,Moult:2019vou,Dixon:2019uzg,Gao:2019ojf,Chen:2020vvp,Ebert:2020sfi}, the development of CFT techniques \cite{Hofman:2008ar,Belitsky:2013xxa,Belitsky:2013bja,Belitsky:2013ofa,Korchemsky:2015ssa,Belitsky:2014zha,Kravchuk:2018htv,Kologlu:2019bco,Kologlu:2019mfz,1822249,Korchemsky:2019nzm}, the application of CFT based techniques to QCD \cite{Chicherin:2020azt,Chen:2020adz,Chen:2021gdk}, and the calculation of higher point correlators \cite{Chen:2019bpb}.

\begin{figure}
\begin{center}
\subfloat[]{
\includegraphics[scale=0.20]{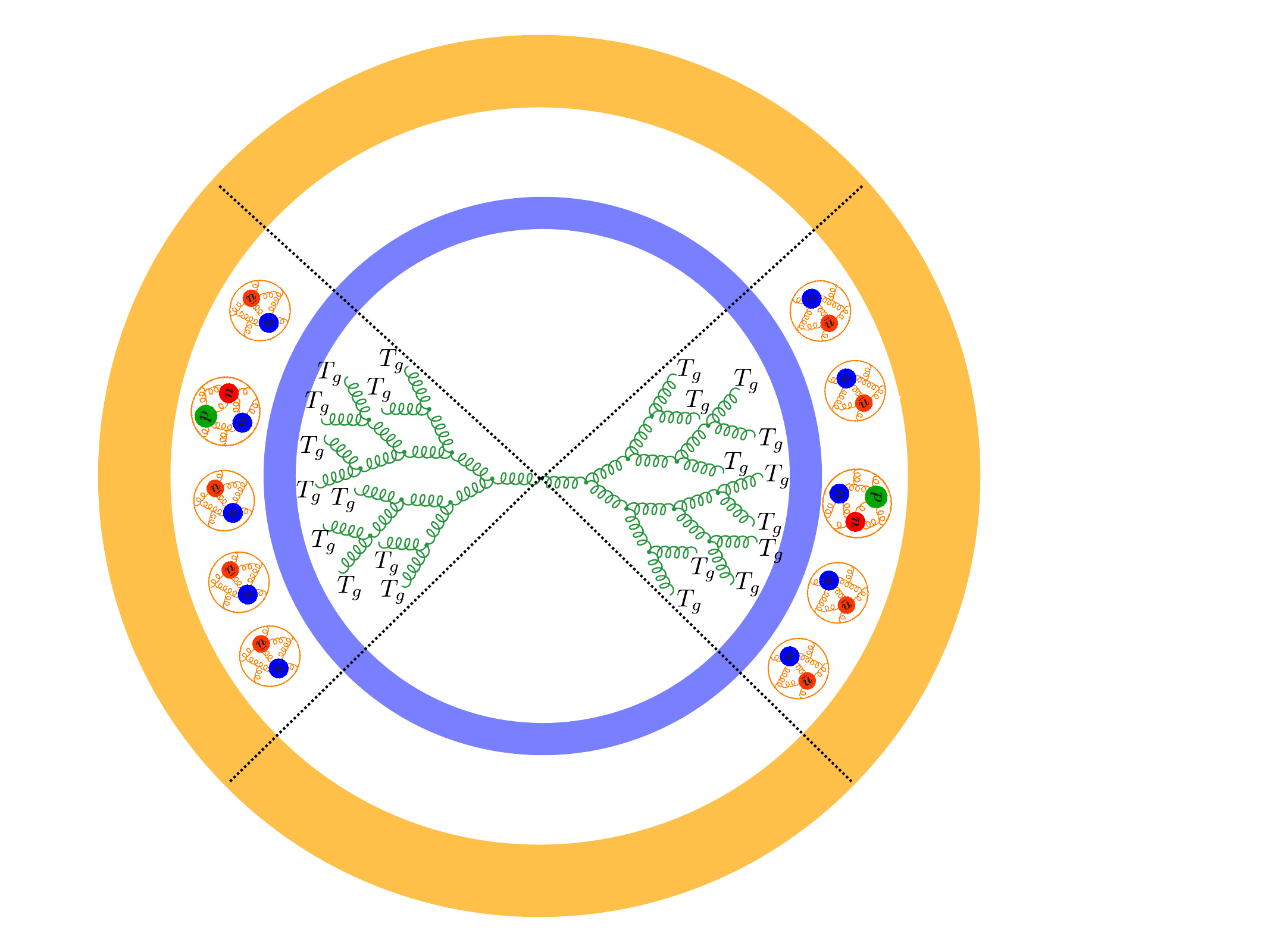}\label{fig:eflow_a}
}
\subfloat[]{
\includegraphics[scale=0.20]{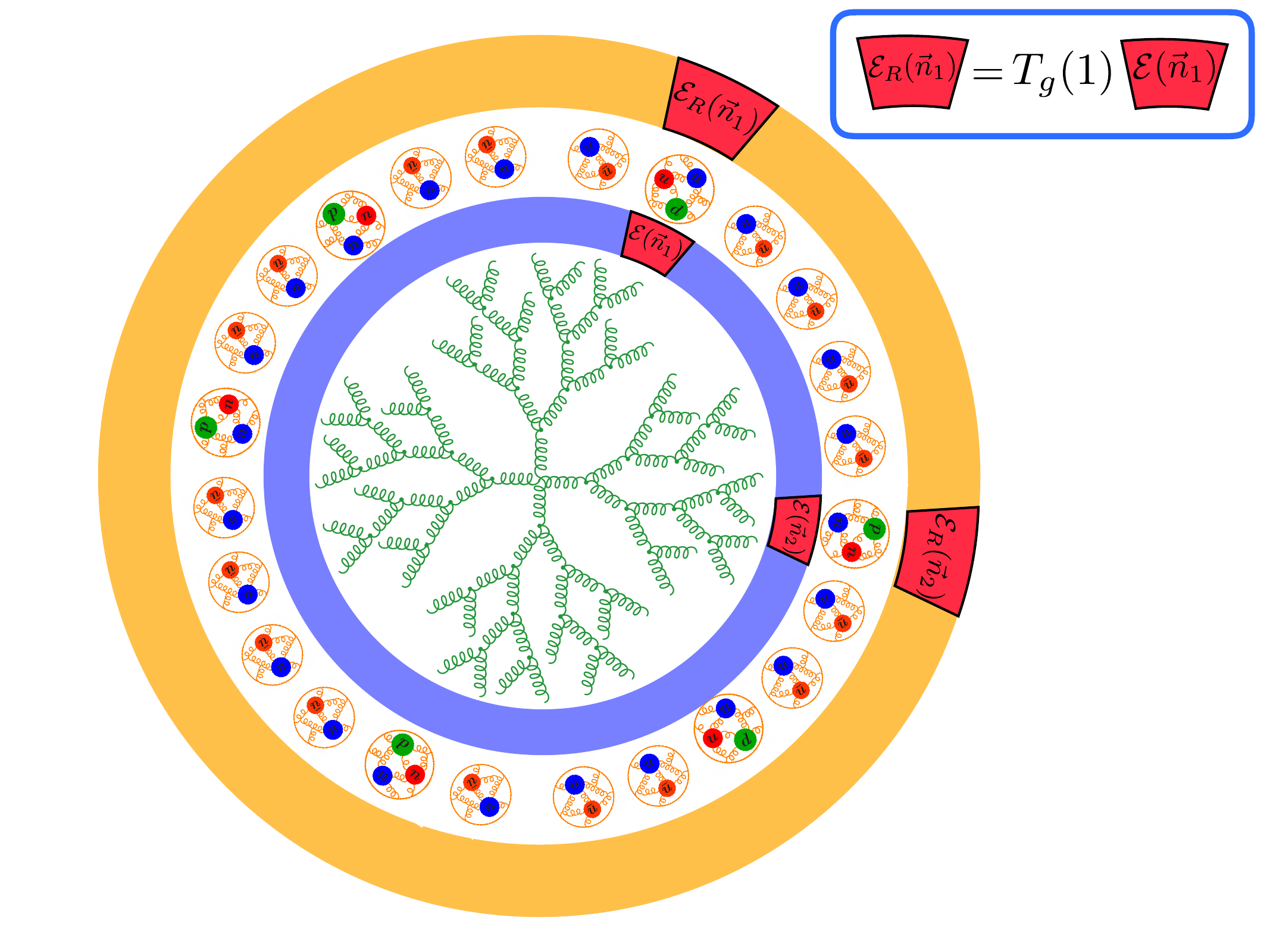}\label{fig:eflow_b}
}\qquad
\end{center}
\caption{(a) For a standard dijet event shape observable, which constrains the phase space of all emissions, a separate track function is needed for every emission, leading to a complicated structure of the hadronization process. (b) For energy correlators, matching can be performed at the level of the detectors, instead of for each parton. Since the number of detectors is fixed this leads to a much simpler  description of the transition from quarks and gluons to hadrons.}
\label{fig:eflow}
\end{figure}

Although these observables appear similar to more standard jet observables, which are typically called ``jet shapes", they are in fact quite different. Jet shapes constrain radiation about some underlying hard process, can be thought of as infrared and collinear safe resolution variables for an $S$-matrix element of quarks and gluons. On the other hand, the correlation functions $\langle \mathcal{E}(\vec n_1) \mathcal{E}(\vec n_2) \cdots \mathcal{E}(\vec n_k) \rangle$, are statistical correlators defined as an ensemble average, and do not constrain the emitted radiation. While these correlators have been well studied in the formal CFT literature, that they can be useful phenomenologically to systematically probe the structure of QCD was emphasized in ref.~\cite{Chen:2020vvp}. 

The energy correlators are simpler perturbatively, which has enabled a number of remarkable calculations in both QCD \cite{Dixon:2018qgp,Luo:2019nig} and $\cN=4$ SYM \cite{Belitsky:2013ofa,Henn:2019gkr}. However, for phenomenological applications to QCD, it is perhaps their non-perturbative simplicity that is even more important, due to the poor current understanding of the hadronization process in QCD. Standard jet or event shape observables are sensitive to the complete structure of emissions. This makes their extension to charged particles (or other subsets $R$ of particles) extremely complicated, since it requires a description of the hadronization process for every single perturbative particle. This is illustrated in \Fig{fig:eflow_a}. Furthermore, in addition to having additional track functions at each perturbative order, the observable also depends on the complete functional form, $T_a(x)$, of these non-perturbative functions. On the other hand, for correlation functions of energy flow operators, the fragmentation process should be thought of as a matching between detector operators in the perturbative and non-perturbative theory. Since the number of detectors is fixed (and in practical applications only low numbers of detectors are considered), this leads to a simple theoretical description of the fragmentation process, that is unchanged order by order in perturbation theory, see \Fig{fig:eflow_b}. It is this simple property of the energy correlators that allows them to be naturally extended to a description of energy flow on subsets of particles. 

 We now formalize this in a factorization theorem involving moments of track functions. This will motivate the study of the renormalization group structure of these moments, which will be the focus of the remainder of this paper. To understand the energy correlators on tracks, we begin by introducing an energy flow operator that only measures energy flow on a restricted set of states, $\cE_R$. This is a fundamentally non-perturbative object, which does not admit a perturbative expansion about free asymptotic quark and gluon states. This restricted energy flow operator admits an OPE onto partonic energy flow operators, 
\begin{align}
\cE_R(\vec n_1)=T_{\bar q}(1) \cE_{\bar q} (\vec n_1)+  T_q(1) \cE_q (\vec n_1)+  T_g(1) \cE_g (\vec n_1) \,.
\end{align}
The matching coefficients are given by first moment of the track function $T_a(1)$, describing the average momentum fraction of the subset $R$, whose formal definition and RG structure will be given in the next section. (Note that track functions can differ between quark flavors, which we ignore here for notational simplicity.)
To study multi-point energy correlators on tracks, one will therefore need to perform the perturbative calculations of the matrix elements
\begin{align}
\langle \cE_{a_1} (\vec n_1) \cE_{a_2} (\vec n_2) \cdots \cE_{a_k} (\vec n_k) \rangle\,.
\end{align}
These are more general than what has been studied in the literature, but the same calculational techniques can be used, as will be discussed in \Sec{sec:results}.

We are now able to present the general form of the factorization formula for a $k$-point correlator in terms of these partonic correlators and moments of track functions 
\begin{align}
\langle \cE_R (\vec n_1) \cE_R (\vec n_2) \cdots \cE_R (\vec n_k) \rangle
&= \sum\limits_{a_1, a_2,\cdots, a_k} T_{a_1}(1)\cdots T_{a_k}(1)  \langle \cE_{a_1} (\vec n_1) \cE_{a_2} (\vec n_2) \cdots \cE_{a_k} (\vec n_k) \rangle \nn \\
&\quad +\text{contact terms}\,.
\end{align}
The contact terms arise when any two detectors are in the same direction,  introducing dependence on higher moments of the track functions.
We will now explicitly show the structure of the contact terms for the two- and three-point correlator. For the two-point correlator, we have
\begin{align}
\langle \cE_R(n_1) \cE_R(n_2) \rangle &=\sum_{a_1,a_2}   T_{a_1}(1) T_{a_2}(1)  \langle \cE_{a_1} (\vec n_1) \cE_{a_2} (\vec n_2) \rangle +\sum\limits_a T_a(2) \langle \cE_a^{(1,1)}(\vec n_1)\rangle \delta(\vec n_1-\vec n_2)\,,
\end{align}
while for the three-point correlator, we have
\begin{align}
\langle \cE_R(n_1) \cE_R(n_2)  \cE_R(n_3)  \rangle &=\sum_{a_1,a_2,a_3}   T_{a_1}(1) T_{a_2}(1) T_{a_3}(1)  \langle \cE_{a_1} (\vec n_1) \cE_{a_2} (\vec n_2)  \cE_{a_3} (\vec n_3) \rangle \nn \\
&\quad +\sum\limits_{a_1,a} T_{a_1}(1) T_a(2) \langle \cE_{a_1}(\vec n_1) \cE_a^{(1,1)}(\vec n_2)\rangle \delta(\vec n_2-\vec n_3)  \nn \\ 
 &\quad +\sum\limits_{a_2,a} T_{a_2}(1) T_a(2) \langle \cE_{a_2}(\vec n_2) \cE_a^{(1,1)}(\vec n_1)\rangle \delta(\vec n_1-\vec n_3) \nn \\
 &\quad +\sum\limits_{a_3,a} T_{a_3}(1) T_a(2) \langle \cE_{a_3}(\vec n_3) \cE_a^{(1,1)}(\vec n_1)\rangle \delta(\vec n_1-\vec n_2) \nn \\
 &\quad +\sum\limits_{a} T_a(3) \langle \cE_a^{(1,1,1)}(\vec n_1)\rangle \delta(\vec n_1-\vec n_2) \delta(\vec n_2-\vec n_3)\,.
\end{align}
The extension to higher point correlators should be clear. These contact terms introduce dependence on higher track function moments $T_a(n)$. The precise operator definition of the corresponding lightray operators, $\cE_a^{(1,1,\cdots, 1)}$, will not be important here, but in perturbation theory these simply weight the state by $E^n$, where $n$ is the number of $1$ in the exponent. The precise notation is chosen due to their relation to multi-hadron fragmentation functions.

One appealing aspect of this factorization formula is that for an $N$-point correlator, it contains a \emph{finite} sum over the different track function structures. This structure is fixed by the properties of the detectors, and independent of the order in perturbation theory, as visualized in \Fig{fig:eflow_b}. This follows the general philosophy arising from CFTs, namely that one should study the space of detectors rather than the states, which leads to significant simplifications here.

\section{Track Function Moments and their Renormalization Group Evolution}\label{sec:GFF}

Having shown how moments of track functions naturally appear in the study of energy flow, in this section we study in detail their renormalization group structure.

\subsection{Definition and Sum Rules}

The track function describes the momentum fraction $x$ of an initial parton $i$ that is converted to a subset $R$ of the final-state hadrons specified in terms of some particular quantum number, e.g.~charge, strangeness, etc. Its definition in terms of a matrix element in quantum field theory is in light-cone gauge given by~\cite{Chang:2013rca,Chang:2013iba}
\begin{align} \label{T_def}
T_q(x)&=\!\int\! \df y^+ \df ^{d-2} y_\perp e^{ik^- y^+/2} \sum_X \delta \biggl( x\!-\!\frac{P_R^-}{k^-}\biggr)  \frac{1}{2N_c}
\text{tr} \biggl[  \frac{\gamma^-}{2} \langle 0| \psi(y^+,0, y_\perp)|X \rangle \langle X|\bar \psi(0) | 0 \rangle \biggr]\,,
 \\
T_g(x)&=\!\int\! \df y^+ \df^{d-2} y_\perp e^{ik^- y^+/2} \sum_X \delta \biggl( x\!-\!\frac{P_R^-}{k^-}\biggr) \frac{-1}{(d\!-\!2)(N_c^2\!-\!1)k^-}
 \langle 0|G^a_{- \lambda}(y^+,0,y_\perp)|X\rangle \langle X|G^{\lambda,a}_- (0)|0\rangle. 
\nn \end{align} 
In general covariant gauges, Wilson lines are required to maintain gauge invariance, as is standard for fragmentation functions. The Fourier transform of $y^+$ fixes the large light-cone momentum of the initiating field to be $k^-$, and the $y_\perp$-integral sets its transverse momentum to zero. The delta function encodes the measurement of the momentum fraction $x$ of the subset $R$ of the final-state $X$. Finally, the matrix elements encode the probability of a quark or gluon to produce a final-state $X$, averaged over its color and spin (with $d$ the number of space-time dimensions, used as regularization). 

We will often work in terms of the moments of the track functions, defined as 
\begin{align} \label{eq:T_mom}
T_a(n,\mu)=\int \limits_0^1 \df x~ x^n~ T_a (x,\mu)\,.
\end{align}
Note that this differs by one unit from the standard convention, which is why the evolution of $T(n,\mu)$ will involve the DGLAP anomalous dimensions $\ga(n+1)$ in the standard convention.
The zeroth moment satisfies the sum rule
\begin{align}\label{T0}
T_a(0,\mu)=1\,,
\end{align}
implying that the track function is normalized. 

\subsection{Comparison to Fragmentation Functions}

The difference between the definition of the track function in \eq{T_def} and the fragmentation function $D_{a \to h}$ is that
\begin{align}\label{D_repl}
   \sum_X \delta\biggl( x-\frac{P_R^-}{k^-}\biggr) |X \rangle \langle X| \quad \longrightarrow \quad
   \int \frac{\df^d p_h}{(2\pi)^{d-1}} \de(p_h^2 - m_h^2) \sum_{X'} \delta\biggl( x-\frac{P_h^-}{k^-}\biggr) |hX' \rangle \langle hX'|   
\,,\end{align}
so instead the momentum fraction $x$ of a hadron $h$ (e.g.~$h=\pi^+$) is measured.

Because a single parton can produce multiple hadrons, the fragmentation function is not normalized, in contrast to \eq{T0}. Instead, it satisfies the momentum sum rule
\begin{align}
  \sum_h D_{a \to h}(1,\mu) = 1
\,,\end{align}
where the sum on $h$ is over all hadron species. Note that this is consistent with \eqs{T0}{D_repl} because
\begin{align}
   \sum_h \int \frac{\df^d p_h}{(2\pi)^{d-1}} \de(p_h^2 - m_h^2) \sum_{X'} \frac{P_h^-}{k^-} |hX' \rangle \langle hX'|  = \sum_{X} |X\rangle \langle X|
\end{align}
In grouping $h$ and $X'$ together in $X$, the factor $P_h^-/k^-$ is necessary to get the correct symmetry factor, because $X'$ may also contain another hadron $h$. This is discussed in sec.~2.5 of ref.~\cite{Jain:2011xz}. 

The first moment of the track function and fragmentation function are related
\begin{align}
  T_a(1,\mu) = \sum_{\text{charged}\ h} D_{a \to h}(1,\mu)
\end{align}
However, for the second moment
\begin{align}
 T_a(2,\mu) = \sum_{\text{charged}\ h} D_{a \to h}(2,\mu) + \sum_{\text{charged}\ h_1,h_2} D_{a \to h_1h_2}(1,1,\mu)
\,,\end{align} 
where $D_{a \to h_1h_2}(1,1,\mu)$ is a moment of the dihadron fragmentation function. This arises because $x = \sum_i x_i$ where $x_i$ is the momentum fractions of the $i$-th hadron in $R$, and $x^2 = \sum_i x_i^2 + \sum_{i \neq j} x_i x_j$. (For the corresponding discussion in the context of jet charge, see ref.~\cite{Waalewijn:2012sv}.) This can be extended to the $n$-th moment of the track function, which involves $n$-hadron fragmentation functions, clearly demonstrating that the track function is sensitive to correlations between final-state hadrons.

\subsection{Renormalization Group Evolution and Shift Symmetries}

The track function evolution has the following general form 
\begin{align}\label{T_evo_x}
  \frac{\df}{\df \ln \mu^2} T_a(x) &= \sum_N \sum_{\{a_f\}} \biggl[  \prod_{i=1}^N \int_0^1\! \df z_i \biggr] \de\Bigl(1 - \sum_{i=1}^N z_i\Bigr) P_{a \to \{a_f\}}(\{z_f\}) 
  \nn\\
 & \quad \times \biggl[\prod_{i=1}^N \int_0^1\! \df x_i\, T_{a_i}(x_i)\biggr] \de\Bigl(x - \sum_{i=1}^N z_i x_i\Bigr) \,,
\end{align}
where we suppressed the argument $\mu$ for brevity. 
There is a sum over all possible splittings of a parton $a$ into partons $a_f$ with momentum fractions $z_f$, and for each of these parton there is a track function $T_{a_i}$. The total momentum fraction $x$ is obtained by summing over the $x_i$ of these partons, which are rescaled because these fractions are with respect to the parton $a_i$ who carry a momentum fraction $z_i$ of the initial parton $a$. The sum on $N$ goes up to the order $\alpha_s^{N-1}$ that one is working to in perturbation theory. E.g.~at order $\alpha_s^2$ we need at most $N=3$, corresponding to $1 \to 3$ collinear splittings. The explicit expression for $P$ is only known at order $\alpha_s$, for which $N=2$.

We note that this evolution equation is invariant when the arguments of all track functions are shifted $T_a(x) \to T_a(x + b)$ and $T_{a_i}(x_i) \to T_{a_i}(x_i + b)$. This follows because $x - \sum_i z_i x_i = (x+b) - \sum_i z_i (x_i+b)$ due to momentum conservation $\sum_i z_i = 1$. Track functions must satisfy $0 \leq x, x_i \leq 1$, and thus for a generic track function this shift cannot physically be performed. However, the evolution equation is independent of the functional form of the track function, so that one can choose to consider a compactly supported track function on which the shift does make physical sense. This allows shifts to be used to constrain the form of the evolution. 

Converting \eq{T_evo_x} to moment space for integer $n$, we can use the multinomial expansion to obtain
\begin{align} \label{T_evo_n}
  \frac{\df}{\df \ln \mu^2} T_a(n) &= \sum_N \sum_{\{a_f\}} \sum_{\{m_f\}} \ga_{a \to \{a_f\}}(\{m_f\})\,  \prod_{i=1}^N T_{a_i}(m_i,\mu) 
  \nn \\
  \ga_{a \to \{a_f\}}(\{m_f\}) &= \begin{pmatrix} & n & \\ m_1 & m_2 & \cdots \end{pmatrix}
  \biggl[\prod_{i=1}^N \int_0^1\! \df z_i\, z_i^{m_i}  \biggr]\de\Bigl(1 - \sum_{i=1}^N z_i\Bigr) P_{a \to \{a_f\}}(\{z_f\})
\,.\end{align}
The sum of the moments of the track functions on the right-hand side must equal $n$, i.e.~$\sum_i m_i = n$. 

The aforementioned shift symmetry of the evolution is particularly convenient for moments:
\begin{align}
T_a(n,\mu)=\int \limits \df x~ x^n~ T_a (x,\mu) \to \int \limits \df x~ x^n~ T_a (x+b,\mu) =  \int \limits \df x~ (x-b)^n~ T_a (x,\mu)
\,.\end{align}
Explicitly, for the first few moments,
\begin{align}
  T_a(0,\mu) \to  T_a(0,\mu) =1\,,
  \quad
  T_a(1,\mu) \to  T_a(1,\mu) - b\,,  
  \quad
  T_a(2,\mu) \to  T_a(2,\mu) - 2 b T_a(1,\mu) + b^2\,.
\end{align}
In the next subsections we will work out the consequences of this, starting with the case of a pure Yang-Mills theory that allows us to ignore flavors.

The evolution of the fragmentation function can be derived from the same $P$ in \eq{T_evo_x}
\begin{align}
  \frac{\df}{\df \ln \mu^2} D_{a \to h}(x) &= \sum_N \sum_{\{a_f\}} \biggl[\prod_{i=1}^N  \int_0^1\! \df z_i \biggr] \de\Bigl(1\! -\! \sum_{i=1}^N z_i\Bigr) P_{a \to \{a_f\}}(\{z_f\}) \sum_{i=1}^N \int_0^1\! \df x_i\, D_{a_i \to h}(x_i) \de(x \!-\! z_i x_i)\,,
\end{align}
In moment space this becomes 
\begin{align} \label{D_evo_n}
  \frac{\df}{\df \ln \mu^2} D_{a \to h}(n) &=- \sum_b  \ga_{ba}(n+1) D_{b\to h}(n)\,,
  \nn \\
  \ga_{ba}(n+1) &=
  -\sum_{N}\sum_{\{a_f\}} \biggl[ \prod_{i=1}^N \int_0^1\! \df z_i\, \biggr] \de\Bigl(1 - \sum_{i=1}^N z_i\Bigr) P_{a \to \{a_f\}}(\{z_f\}) \sum_{i=1}^N \de_{b, a_i} z_i^n 
\,.\end{align}
Here we have used the standard conventions for the timelike twist-two spin-$n$, anomalous dimensions, $\gamma(n)$.
A comparison of \eqs{T_evo_n}{D_evo_n} reveals that the coefficient of the anomalous dimension of $T_a(n)$ involving $T_b(n)$ is the same as that entering in the evolution of the moment $D_{a \to h}(n)$ of the fragmentation function, 
\begin{align}
 -\ga_{ba}(n+1) &= \sum_N\sum_{ \{a_f\} } \Big( \ga_{a \to \{a_f\}}(\{n,0,\cdots,0\}) \de_{b,a_1} \!+\! \ga_{a \to \{a_f\}}(\{0,n,\cdots,0\}) \de_{b,a_2} \nn \\
 & \!+\!   \cdots \!+\! \ga_{a \to \{a_f\}}(\{0,0,\cdots,n\}) \de_{b,a_N} \Big)
\,.\end{align}

\subsection{Constraints from Shift Symmetry: Pure Yang-Mills theory}\label{sec:glu}

We will now demonstrate how the shift-symmetry determines the structure of the evolution equation for a pure Yang-Mills theory.\footnote{Note that in this case the electric charge is not relevant, but one could use track functions to describe the momentum fraction of bound states of e.g.~a specific type of glueball.}
From the form of \eq{T_evo_n} we know that 
\begin{align} 
    \frac{\df}{\df \ln \mu^2} T(1) = \ga_1 T(1)\,, 
    \quad
    \frac{\df}{\df \ln \mu^2} T(2) = \ga_2 T(2) + \ga_{11} T(1)^2\,,
\end{align}
etc. Since we have only a gluon, we suppress flavor labels. The notation $\ga_1, \ga_2, \ga_{11}, \dots$ for the anomalous dimensions is only used in the pure gluon case described here and in app.~\ref{app:yang_mills}. From the perspective of the shift symmetry alone, these anomalous dimensions are arbitrary. We will later relate them to the timelike twist-2 spin-$n$ anomalous dimensions $\gamma(n)$ (note the differing notation).

 Applying the shift to these equations, we obtain
\begin{align}
    \frac{\df}{\df \ln \mu^2} (T(1) - b) &= \ga_1 (T(1) - b)\,, 
    \nn \\
    \frac{\df}{\df \ln \mu^2} (T(2) - 2b T(1) + b^2) &= \ga_2 (T(2)  - 2b T(1) + b^2)+ \ga_{11} (T(1)-b)^2
\,,\end{align}
which leads to 
\begin{align}
    \frac{\df}{\df \ln \mu^2} T(1) =  \ga_1 T(1) - \ga_1 b\,,
\end{align}
and thus $\ga_1 = 0$ in this case (this is not true when there are other parton species), as well as 
\begin{align} \label{g_evo_2}
    \frac{\df}{\df \ln \mu^2} T(2) = \ga_2 T(2) + \ga_{11} T(1)^2   + (\ga_{11} + \ga_2) (2b T(1) + b^2)
\,,\end{align}
implying $\ga_{11} = - \ga_{2}$.

A more economical approach to deriving these equations is to directly use \emph{shift-invariant} central moments
\begin{align}
\label{eq:centralmoment}
\sigma(n,\mu)=\int \limits_0^1 \df x~ (x - \langle x \rangle)^n~ T(x,\mu)\,,
\end{align}
where the average $\langle x \rangle$ is simply the first moment $T(1,\mu)$. Note that this can simply be thought of as a change of basis. Now we immediately have
\begin{align}
    \frac{\df}{\df \ln \mu^2} \sigma(2) = \ga_2\, \sigma(2)
\,,\end{align}
since no other terms can appear on the right-hand side. Inserting $\sigma(2) = T(2) - T(1)^2$, we then again obtain $\ga_{11} = - \ga_{2}$. As we will see, in the case of multiple flavors one can form shift invariant first moments, $T_i(1) - T_j(1)$.

Extending this to higher moments, we obtain the general structure of the renormalization group evolution of the central moments of the track functions
\begin{align}
    \frac{\df}{\df \ln \mu^2} \sigma(3) &= \ga_3\, \sigma(3) \,,\nn\\
    \frac{\df}{\df \ln \mu^2} \sigma(4) &= \ga_4\, \sigma(4) + \ga_{22} \sigma(2)^2 \,,\nn \\
    \frac{\df}{\df \ln \mu^2} \sigma(5) &= \ga_5\, \sigma(5) + \ga_{32} \sigma(3) \sigma(2) \,,\nn \\
    \frac{\df}{\df \ln \mu^2} \sigma(6) &= \ga_6\, \sigma(6) + \ga_{42} \sigma(4) \sigma(2) + \ga_{33} \sigma(3)^2 + \ga_{222} \sigma(2)^3 \,,\nn \\
    \frac{\df}{\df \ln \mu^2} \sigma(7) &= \ga_7\, \sigma(7) + \ga_{52} \sigma(5) \sigma(2) + \ga_{43} \sigma(4) \sigma(3) + \ga_{322} \sigma(3) \sigma(2)^2    
\,,\nn \\    
    \frac{\df}{\df \ln \mu^2} \sigma(8) &= \ga_8\, \sigma(8) + \ga_{62} \sigma(6) \sigma(2) + \ga_{53} \sigma(5) \sigma(3) + \ga_{44} \sigma(4)^2 + \ga_{422} \sigma(4) \sigma(2)^2  + \ga_{332} \sigma(3)^2 \sigma(2)
\,,\nn \\    
    \frac{\df}{\df \ln \mu^2} \sigma(9) &= \ga_9\, \sigma(9) + \ga_{72} \sigma(7) \sigma(2) + \ga_{63} \sigma(6) \sigma(3) + \ga_{54} \sigma(5)\sigma(4) + \ga_{522} \sigma(5) \sigma(2)^2 
    \nn \\ & \quad
     + \ga_{432} \sigma(4) \sigma(3) \sigma(2) + \ga_{333} \sigma(3)^3 
\,,\end{align}
etc. Because the evolution of $T(n)$ can involve at most 3 track functions at order $\alpha_s^2$, the form of these equations are further restricted at this order. Thus, \emph{up to order} $\alpha_s^2$,
\begin{align} \label{eq:ga_rel}
 \ga_{22} &= 6 \ga_2 - 8 \ga_3 + 3 \ga_4\,, \\
 \ga_{32} &= 10 \ga_2 -10 \ga_3 + 2 \ga_5\,, \nn \\
 \ga_{222} &= -\ga_{42} +15 \ga_2 - 40 \ga_3 +60 \ga_4 - 48 \ga_5 +15 \ga_6\,, \nn \\
 \ga_{33} &= -\ga_{42} +15 \ga_2 - 20 \ga_3 + 15 \ga_4 - 12 \ga_5 + 5 \ga_6\,, \nn \\ 
  \ga_{52} &= \tfrac73 \ga_{42} -14 \ga_2 + \tfrac{70}3 \ga_3 -35 \ga_4 + 49 \ga_5 - 35 \ga_6 + 9 \ga_7\,, \nn \\ 
  \ga_{43} &= -\tfrac73 \ga_{42} + 35 \ga_2 - \tfrac{175}3 \ga_3 + 70 \ga_4 - 70 \ga_5 + 35 \ga_6 - 5 \ga_7\,, \nn \\   
  \ga_{322} &= -\tfrac73 \ga_{42} + 35 \ga_2 - \tfrac{280}3 \ga_3 + 140 \ga_4 - 112 \ga_5 + 35 \ga_6
\,, \nn \\  
 \ga_{53} &=\tfrac{28}{3} \ga_{42} - 3 \ga_{62} - 56 \ga_2 + \tfrac{280}{3} \ga_3 - 140 \ga_4 + 168 \ga_5 - 56 \ga_6 - 48 \ga_7 + 28 \ga_8
 \,, \nn \\
 \ga_{44} &= -\tfrac{28}{3} \ga_{42} + 2 \ga_{62} +84 \ga_2 - \tfrac{448}{3} \ga_3 +210 \ga_4 - 224 \ga_5 + 84 \ga_6 + 32 \ga_7 - 21 \ga_8\,, \nn \\
 \ga_{422} &= \tfrac{28}{3} \ga_{42} - 3 \ga_{62} - 56 \ga_2 + \tfrac{112}{3} \ga_3 + 70 \ga_4 - 224 \ga_5 + 364 \ga_6 - 288 \ga_7 + 84 \ga_8
 \,,\nn \\
 \ga_{332} &= -\tfrac{28}{3} \ga_{42} + 2 \ga_{62}+84 \ga_2 - \tfrac{448}{3} \ga_3 + 140 \ga_4 - 196 \ga_6 + 192 \ga_7 - 56 \ga_8
 \,, \nn \\
 \ga_{63} &= \tfrac92 \ga_{62} - \tfrac72 \ga_{72} - 42 \ga_6 + 126 \ga_7 - 126 \ga_8 +42 \ga_9 \,, \nn \\
 \ga_{54} &= - \tfrac92 \ga_{62} + \tfrac52 \ga_{72} + 36 \ga_2 - 84 \ga_3 + 126 \ga_4 - 126 \ga_5 + 126 \ga_6 - 162 \ga_7 + 126 \ga_8 - 36 \ga_9 \,, \nn \\
 \ga_{522} &= \!-\! 84 \ga_{42} \!+\! \tfrac{81}{2} \ga_{62} \!-\! \tfrac{27}{2} \ga_{72} \!+\! 612 \ga_{2} \!-\! 1092 \ga_3 \!+\! 1638 \ga_4 \!-\! 1764 \ga_5 \!+\! 126 \ga_6 \!+\!1458 \ga_7 \!-\! 1134 \ga_8 \!+\! 270 \ga_9 \,, \nn \\
 \ga_{432} &= 168 \ga_{42} \!-\!\tfrac{153}{2} \ga_{62}\!+\!\tfrac{45}{2} \ga_{72} \!-\! 1188 \ga_2 \!+\! 1932 \ga_3 \!-\! 2520 \ga_4 \!+\! 2268 \ga_5 \!+\! 882 \ga_6 \!-\! 3294 \ga_7 \!+\! 2142 \ga_8 \!-\! 450 \ga_9\,, \nn \\
 \ga_{333} &= -84 \ga_{42} \!+\! 36 \ga_{62} \!-\! 10 \ga_{72} \!+\! 612 \ga_2 \!-\! 1008 \ga_3 \!+\! 1260 \ga_4 \!-\! 1008 \ga_5 \!-\! 588 \ga_6 \!+\! 1656 \ga_7 \!-\! 1008 \ga_8 \!+\! 200 \ga_9 
\,.\nn \end{align}
This structure for the evolution is fixed \emph{entirely} by shift symmetry alone. However, this does not fix the values of the anomalous dimensions.  To further fix the anomalous dimensions, we note that from their definition, the diagonal anomalous dimensions $\ga_n$ are related to the timelike twist-2 anomalous dimensions (moments of the gluon fragmentation function), $\gamma_{gg}(n)$, by 
\begin{align}
\gamma_n=-\gamma_{gg}(n+1)\,.    
\end{align}
These anomalous dimensions are known to NNLO \cite{Chen:2020uvt,Mitov:2006ic,Moch:2007tx,Almasy:2011eq}. 

Therefore up to $\sigma_5$ all anomalous dimensions are constrained in terms of the DGLAP splitting functions, for $\sigma_6$ only one new anomalous dimension needs to be calculated and no new one is needed for $\sigma_7$. Beyond $\sigma_7$, one (or more) new anomalous dimensions need to be calculated for every moment.

An alternate approach is to exploit the symmetry of the matrix elements. This is in practice equivalent to the shift symmetry, though restricted to a specific order in perturbation theory. For example, at order $\alpha_s^2$ for which $N=3$, we can express the $\ga$ in the equations above to that in \eq{T_evo_n},
\begin{align}
 \ga_0 = \ga(0,0,0) = 0\,,\qquad
 \ga_1 = \ga(1,0,0) + \ga(0,1,0) + \ga(0,0,1) = \ga(0,0,0) = 0
\,,\end{align}
using momentum conservation $z_1 + z_2 + z_3 = 1$. Similarly,
\begin{align}
 \ga_2 &= \ga(2,0,0) + \ga(0,2,0) + \ga(0,0,2) = 3 \ga(2,0,0) \,,\\
 \ga_{11} &=  \ga(1,1,0) + \ga(1,0,1) + \ga(0,1,1) = 3 \ga(1,1,0) = 3 (\ga(0,0,0) - 2\ga(1,0,0) - \ga(2,0,0)) = -\ga_2
\,,\nn\end{align}
using the symmetry under permutations of $z_1, z_2, z_3$.
In the final steps we used that  \emph{under the integral} the following identities hold
\begin{align}
  2z_1 z_2 = (z_1+z_2)^2 - z_1^2 - z_2^2 = (1-z_3)^2 - z_1^2 - z_2^2 = 1 - 2z_3 + z_3^2 - z_1^2 - z_2^2 = 1 - 2z_1 - z_1^2
\,.\end{align}
Clearly the use of shift-symmetric central moments is much simpler.

\subsection{Constraints from Shift Symmetry: Multi-Flavor} \label{sec:constr_multi}

Having described in detail how shift symmetry constrains the form of the evolution in the case of a pure gluon theory, we here extend the discussion to the case of multiple parton species, which is needed for QCD. We will consider the case of one quark species and assume that the track functions for quarks and anti-quarks are the same, to keep the discussion simple and highlight the new features. The extension to multiple quarks is straightforward, and our final results do not use this assumption. 

The simplifying feature of the pure gluon evolution is that the mean, $T(1)$, is not shift invariant, and therefore cannot appear in the evolution equations. Shift symmetry, combined with the uniqueness of the shift invariant second and third moments, then fixes to all orders in perturbation theory the evolution equations for the second and third moments
\begin{align}
    \frac{\df}{\df \ln \mu^2} \sigma(2) &= -\gamma(3)\, \sigma(2)\,, \nn \\
        \frac{\df}{\df \ln \mu^2} \sigma(3) &=- \gamma(4)\, \sigma(3) 
\,.\end{align}
When moving to multiple flavors there are two new features that appear. The first is a trivial extension, namely that we must extend the evolution equations to be matrix equations in flavor space, as is familiar from DGLAP. Focusing for simplicity on the case of one quark and one gluon, we define
\begin{align}
\vec \sigma(n)=
  \begin{pmatrix}
    \sigma^g(n)
\\
\sigma^q(n)
  \end{pmatrix}\,,
\end{align}
as well as the standard matrix of anomalous dimensions
\begin{align}
\hat \gamma(n)=
  \begin{pmatrix}
    \gamma_{gg}(n) &  \gamma_{qg}(n)
\\
   \gamma_{gq}(n) & \gamma_{qq}(n)
  \end{pmatrix} \,.   
\end{align}
The second extension that appears in the case of multiple flavors is a more non-trivial modification, namely the appearance of a new shift invariant quantity, 
\begin{align}
\Delta=T_q(1)-T_g(1)\,,
\end{align}
constructed from the difference of first moments. This object can appear in the evolution equations, leading to additional complexity. 

Focusing on the first five moments, which makes the general structure clear, shift invariance then implies that to all orders in perturbation theory, 
\begin{align}\label{eq:structure}
 \frac{\df}{\df\ln\mu^2}\Delta&= -(\gamma_{qq}(2)+\gamma_{gg}(2))  \Delta\,, \\
 \frac{\df}{\df\ln\mu^2} \vec \sigma(2) &=-\hat \gamma(3) \vec \sigma(2) + \vec \gamma_{\Delta^2}  \Delta^2\,, \nn\\
 \frac{\df}{\df\ln\mu^2} \vec \sigma(3) &=-\hat \gamma(4) \vec \sigma(3) + \hat \gamma_{\sigma_2 \Delta} \vec \sigma(2) \Delta+ \vec \gamma_{\Delta^3} \Delta^3\,, \nn\\
 \frac{\df}{\df\ln\mu^2} \vec \sigma(4) &=-\hat \gamma(5) \vec \sigma(4)+ \hat \gamma_{\sigma_2 \sigma_2}( \vec \sigma(2)\cdot\vec \sigma(2)^T) +\hat \gamma_{\sigma_3 \Delta} \vec \sigma(3) \Delta+ \hat \gamma_{\sigma_2 \Delta^2} \vec \sigma(2) \Delta^2+ \vec \gamma_{\Delta^4} \Delta^4\,, \nn \\
 \frac{\df}{\df\ln\mu^2} \vec \sigma(5) &=-\hat \gamma(6) \vec \sigma(5)+ \hat \gamma_{\sigma_3 \sigma_2}( \vec \sigma(3)\cdot\vec \sigma(2)^T)\nn \\
& \quad + \hat \gamma_{\sigma_4 \Delta} \vec \sigma(4) \Delta+ \hat \gamma_{\sigma_2^2 \Delta} (\vec \sigma(2)\cdot \vec \sigma(2)^T) \Delta+ \hat \gamma_{\sigma_3 \Delta^2} \vec \sigma(3) \Delta^2+ \hat \gamma_{\sigma_2 \Delta^3} \vec \sigma(2) \Delta^3+ \vec \gamma_{\Delta^5} \Delta^5\,.
\nn\end{align}
The presence of $\Delta$ significantly complicates the form of the evolution compared with the pure gluon case, and in particular, the first three moments are no longer uniquely fixed by the shift symmetry. Note that the anomalous dimensions $\hat \gamma_{\sigma_2 \sigma_2}$, $\hat \gamma_{\sigma_3 \sigma_2}$ and $\hat \gamma_{\sigma_2^2 \Delta}$ are rank 3 tensors, taking a matrix as input and returning a vector.

The additional complexity arising from the presence of quarks can be thought of in the two different ways discussed in \Sec{sec:glu}: From the shift-symmetry perspective, the complexity arises purely from the presence of the new invariant $\Delta$. From the perspective of the calculation from matrix elements (discussed briefly at the end of \Sec{sec:glu} and made more concrete in sec.~\ref{sec:calc_split}), the presence of quarks implies that one can no longer symmetrize over the final state particles when using momentum conservation arguments to reduce integrals. The differences that arise from this lack of ability to symmetrize are then captured by powers of $\Delta$. The integrals for these residual $\Delta$-dependent pieces turn out to be simpler to compute.

Despite the fact that the terms proportional to $\Delta$ are not fixed in terms of the DGLAP kernels, we will see that this organization still proves extremely useful, particularly for the case of track functions describing the momentum fraction of charged particles in QCD. In the high energy limit, where the energy cost to produce pions is negligible, one expects that the average properties of the track functions are fixed by isospin, namely $T_g(1)\simeq T_q(1)\simeq 2/3$,  and $\Delta \simeq 0$. This intuition is born out by the evolution equation for $\Delta$ in \Eq{eq:structure}, where the positivity of $\gamma_{qq}(2)+\gamma_{gg}(2)$ drives $\Delta\to0$ at asymptotic energies. This behavior is already well born out at moderate energies, where one finds the approximate numerical relation $\Delta^2/\sigma_2 \sim a_s^{3/2}$, showing that its contribution to the evolution of the second moment is suppressed in the perturbative expansion of the evolution. We will show in \Sec{sec:delta}, the NLO terms proportional to $\Delta$ in the evolution of the second moment are irrelevant even compared to the NNLO DGLAP corrections. For the third moment, the corrections in $\Delta$ are effectively suppressed by one order in the perturbative expansion. This allows us to extend our results for the first three moments to NNLO, which is the most important practical application of the shift symmetry.

The shift symmetry also forces the evolution of the first moments to be proportional to $\Delta$, namely
\begin{align}
 \frac{\df}{\df\ln\mu^2} T_q(1)&=-\gamma_{qq}(2) \Delta\,, \\    
\frac{\df}{\df\ln\mu^2} T_g(1)&=-\gamma_{qg}(2) \Delta\,.
\end{align}
This result also follows from energy conservation in the one point function $\langle \cE(\vec n_1) \rangle$, further emphasizing the connection between the shift symmetry and energy conservation. 
This result shows that the evolution of the first moments of the track functions is numerically suppressed by a factor of $\Delta/T(1)$, as compared to the naive expectation. The inclusion of tracks in factorization formulas for energy correlators will therefore have an extremely minor effect, explaining the observation of \cite{Chen:2020vvp}. 

Finally, one appealing feature of the structure of the equations in \Eq{eq:structure} is that it is known that the eigenvalues of the $\hat\gamma(N)$ are positive. This allows us to immediately see that the cumulants (or central moments) of the track functions decay to zero. In the high energy limit, they converge to a delta function with $\Delta=0$, which is the unique attractive fixed point of the evolution. The limiting value of $T_q(1) = T_g(1)$, corresponding to the position of the delta function, is the only nonperturbative parameter that remains.

\section{Track Function Moments at NLO}\label{sec:results}

Having discussed the general structure of the RG evolution of track function moments in \Sec{sec:GFF},  we now move on to their calculation in QCD. We describe our calculational technique in \Sec{sec:calc}, and present the full results for the first six moments in \Sec{sec:res_explicit}. For simplicity, throughout this section we use the language of track functions for charged particles, as opposed to a generic subset of particles. However, our calculations are completely generic, and can be applied to any general subset, $R$, of hadrons.

\subsection{Calculational Technique}\label{sec:calc}

To verify the universality of the renormalization of the moments of the track functions, we compute it in two different ways: First we use an IRC safe observable that is directly sensitive to the track function moments, namely the EEC and projected EECs. When computed on tracks, this observable is no longer IRC safe, and the infrared poles directly determine the RG evolution of the track function moments. Second, we compute the moments of the track function by computing a jet function on tracks. This approach is computationally much simpler since it only requires the integration of splitting functions instead of complete matrix elements, but it assumes collinear factorization, and hence the universality of the track functions. The agreement between these two approaches provides a strong check both on our calculations and on the universality of the track functions. The universality of the first three moments of the track functions was tested at NLO in this same manner in \cite{Li:2021zcf}. Here we extend this to the sixth moment. In the following two subsections we detail these two approaches.

\subsubsection{Using Projected Energy Correlators}\label{sec:calc_correlator}

We begin by computing the RG for the track functions from the structure of infrared poles in energy-energy correlators, which was briefly described in \cite{Li:2021zcf} for the case of the two-point correlator. Here we describe it in some detail, as well as its extension to projected energy correlators, which is necessary to extract the RG of higher moments of the track functions.

The standard two-point energy correlator~\cite{Basham:1978bw,Basham:1978zq,Richards:1982te} is defined as 
\begin{align}
  \label{eq:EECdef}
  \frac{\df \sigma}{\df z}= \sum_{i,j}\int \df \sigma\ \frac{E_i E_j}{Q^2} \delta\Bigl(z - \frac{1 - \cos\chi_{ij}}{2}\Bigr) \,.
\end{align}
This can be extended to a projected $N$-point energy correlator~\cite{Chen:2020vvp}, which is sensitive to higher point correlations, but is only differential in the longest side, $z_L$. It is defined as 
\begin{align}
  \label{eq:projected_mom}
  \frac{\df \sigma^{[N]}}{\df z_L} & = \sum_m\sum_{1\leq i_1,\ldots, i_N \leq m} 
\int\! \df \sigma_{e^+e^- \to X_m}
\frac{\prod_{j=1}^N E_{i_j}}{Q^N}
\, \delta(z_L - \max \{z_{i_1 i_2}, z_{i_1 i_3}, \ldots , z_{i_{N-1} i_N } \}) \,,
\end{align}
where $X_m$ denotes a $m$-particle final state and 
$z_{ij} = (1 - \vec{n}_i \cdot \vec{n}_j)/2 = (1 -  \cos\theta_{ij})/2$ is the two-particle angular distance. 

The projected correlators are IRC safe observables. However, when computed on tracks, they have collinear divergences. These collinear divergences must be absorbed by the track functions. Therefore by computing these collinear divergences, we can obtain the RG of the track functions. To simplify the notation, we combine all the products of track functions of a fixed total weight $n$ (see \eqref{T_evo_n})  into a vector $\vec {\bf T}_n$ (e.g.~for $n=2$, $\vec {\bf T}_2=\{T_g(2),T_q(2), T_q(1)T_q(1), T_g(1)T_q(1),$ $T_g(1) T_g(1) \}$). For notational simplicity, throughout this section we consider the case of a single flavor of quarks, and make the assumption $T_q=T_{\bar q}$. However, we have performed the complete calculation without this assumption.
Writing the renormalization group evolution of $\vec {\bf T}_n$ as
\begin{align}
\frac{\df}{\df\ln \mu^2}\vec{ \bf T}_n=\widehat R_n~\vec {\bf T}_n\,,
\end{align}
then 
\begin{align}\label{eq:poles}
\vec {\bf T}_{n,\text{bare}}&=\vec {\bf{T}}_n(\mu)+a_s \frac{\widehat{R}_n^{(1)}}{\epsilon} \vec{\bf{T}}_n(\mu)
+ \frac{1}{2} a^2_s\left( \frac{\widehat{R}_n^{(2)}}{\epsilon}+\frac{ \widehat{R}_n^{(1)}\widehat{R}_n^{(1)}-\beta_0 \widehat{R}_n^{(1)}} {\epsilon^2}  \right) \vec{\bf{T}}_n(\mu)+\mathcal{O}(a_s^3)\,,\\
&\equiv \widehat{\Gamma}_n(a_s,\epsilon)\vec {\bf{T}}_n(\mu)\, .\nn
\end{align}
where $a_s = \alpha_s(\mu)/(4\pi)$.

In terms of the tree-level track functions $T^{(0)}$, we can write the two-point track EEC as
\begin{align} \label{sigma_T_bare_v2}
\left(\frac{\df \Sigma}{\df z}\right)_\text{tr}&=\sum_{a,b\in \{q_j,\bar{q}_j,g\} }T^{(0)}_{a}(1)T^{(0)}_{b}(1)\ \frac{\df\Sigma_{ab}}{\df z}+\sum_{c\in \{q_j,\bar{q}_j,g\} } T^{(0)}_c(2)\ \frac{\df\Sigma_{c^2}}{\df z} \,.
\end{align}
The perturbatively calculable components entering this formula are given by
\begin{align} \label{eq:sigma_T_bare}
\frac{\df\Sigma_{ab}}{\df z}&=\sum_m \sum_{1 \leq i_1\neq i_2 \leq m} \int \df \Phi_m |\mathcal{M}_m|^2\, \de_{a,f_{i_1}} \de_{b,f_{i_2} } \frac{E_{i_1}E_{i_2}}{Q^2} \delta\Bigl(z-\frac{1-\cos\chi_{i_1i_2}}{2}\Bigr)  \,, \nn \\
\frac{\df\Sigma_{c^2}}{\df z}&=\sum_m \sum_{1 \leq i \leq m} \int \df \Phi_m |\mathcal{M}_m|^2\, \de_{c,f_i} \frac{E_{i}^2}{Q^2}\delta(z)\,.
\end{align}
Here $f_{i_1},f_{i_2},f_i$ denote the flavors of the final-state partons with the four-momenta $p^\mu_{i_1},p^\mu_{i_2},p^\mu_{i}$, $\de_{a,i_1}$, $\de_{b,i_2}$ and $\de_{c,i}$ are Kronecker deltas in flavor space, $\df \Phi_m$ denotes $m$-body phase space and $\mathcal{M}_m$ is the corresponding matrix element.

Using that in dimensional regularization the loop corrections to the track function are scaleless, $T^{(0)} = T_{\rm bare}$, we can employ \eqref{eq:poles} to rewrite \eqref{eq:sigma_T_bare}  in terms of the renormalized track functions,
\begin{align}
\left(\frac{\df \Sigma}{\df z}\right)_\text{tr}&=\!\!
&
\frac{\df \vec{\Sigma}}{\df z} \cdot 
\begin{matrix}
\underbrace{
\begin{matrix}
\widehat{\Gamma}_2\\
\overbrace{
\left[
\mathbf{1}+a_s\frac{\widehat{R}_2^{(1)}}{\epsilon}+\frac{1}{2}a_s^2\left(\frac{\widehat{R}_2^{(2)}}{\epsilon}+\frac{\widehat{R}_2^{(1)}\widehat{R}_2^{(1)}-\beta_0\widehat{R}_2^{(1)}}{\epsilon^2}\right)+\mathcal{O}(a_s^3)
\right]
}
\end{matrix}
\begin{matrix}
\vec{\mathbf{T}}_2(\mu)\\
\overbrace{\begin{pmatrix}
T_g(2)\\
T_{q_1}(2)\\
\cdots\\
T_{q_{n_f-1}}(1)T_{q_{n_f}}{(1)}
\end{pmatrix}}\\
\phantom{a}
\end{matrix}
}\\
\vec{\mathbf{T}}_{2,\text{bare}}
\end{matrix}\,.
\end{align}
The UV poles of the track function renormalization must cancel against the IR poles in $\vec \Sigma$ to yield a finite result, allows us to extract the RG evolution of the first and second moments of the track function.

To have access to the higher moments of the track functions, we must consider the higher point projected correlators. These proceed in a similar manner.
Focusing on the three-point projected correlators, we have 
\begin{align}
\left(\frac{\df \Sigma}{\df z_L}\right)_\text{tr}&=\sum_{a,b,c\in \{q_j,\bar{q}_j,g\}} \frac{\df\Sigma_{abc}}{\df z_L}  T^{(0)}_{a}{(1)}T^{(0)}_{b}{(1)} T^{(0)}_{c}{(1)}   +
\sum_{a,b\in \{q_j,\bar{q}_j,g\}} \frac{\df\Sigma_{ab^2}}{\df z_L}  T^{(0)}_{a}{(1)}T^{(0)}_{b}{(2)}  
\nn \\ & \quad
+ \sum_{c\in \{q_j,\bar{q}_j,g\}} \frac{\df\Sigma_{c^3}}{\df z_L} T^{(0)}_{c}{(3)}\,.
\end{align}
The perturbatively calculable components entering this formula are
\begin{align} \label{eq:sigma_T_bare}
\frac{\df\Sigma_{abc}}{\df z_L}&=\sum_m \sum_{1 \leq i_1\neq i_2\neq i_3 \leq m} \int \df \Phi_m |\mathcal{M}_m|^2\, \de_{a,f_{i_1}} \de_{b,f_{i_2}} \de_{c,f_{i_3}} \frac{E_{i_1}E_{i_2}E_{i_3}}{Q^3} \delta\Bigl(z_L-\frac{1-\cos\chi_{L}}{2}\Bigr)  \,, \nn \\ 
\frac{\df\Sigma_{ab^2}}{\df z_L}&=\sum_m \sum_{1 \leq i_1\neq i_2 \leq m} \int \df \Phi_m |\mathcal{M}_m|^2\, \de_{a,f_{i_1}} \de_{b,f_{i_2} } \frac{E_{i_1}E_{i_2}^2}{Q^3} \delta\Bigl(z_L-\frac{1-\cos\chi_{i_1i_2}}{2}\Bigr)  \,,\nn \\
\frac{\df\Sigma_{c^3}}{\df z_L}&=\sum_m \sum_{1 \leq i \leq m} \int \df \Phi_m |\mathcal{M}_m|^2\, \de_{c,f_i} \frac{E_{i}^3}{Q^3}\delta(z_L)\,.
 \end{align}
These have the same structure as for the two-point correlator, with the only difference being the higher energy weights. They can therefore be computed using the same techniques. The integrals $\Sigma_{abc}$ are more complicated, but fortunately the shift symmetry can be used to reconstruct the full answer from just $\Sigma_{ab^2}$ and $\Sigma_{c^3}$ (at least to the order at which we are currently working).
More generally, for the evolution of the higher moments of the track functions, we consider the integrals 
\begin{align} 
\frac{\df \Sigma_{a^pb^q}}{\df z_L}&=\sum_m \sum_{1 \leq i_1\neq i_2 \leq m} \int \df \Phi_m |\mathcal{M}_m|^2\, \de_{a,i_1} \de_{b,i_2} \frac{E_{i_1}^pE_{i_2}^q}{Q^{p+q}} \delta\Bigl(z_L -\frac{1-\cos\chi_{i_1i_2}}{2}\Bigr)  \,, \nn \\
\frac{\df\Sigma_{c^p}}{\df z_L}&=\sum_m \sum_{1 \leq i \leq m} \int \df \Phi_m |\mathcal{M}_m|^2\, \de_{c,i} \frac{E_i^p}{Q^p}\delta(z_L)\,.
\end{align}
and then use the shift symmetry to reconstruct the full result.

These integrals can be computed using the same approach as was used to compute the standard energy correlator in ref.~\cite{Dixon:2018qgp}, and subsequently in refs.~\cite{Luo:2019nig,Gao:2020vyx}. This approach is an extension of the reverse unitarity method \cite{Anastasiou:2002yz}, which expresses delta functions from phase space constraints in terms of propagators allowing more standard loop integration techniques to be used.  Using the Cutkosky rules \cite{Cutkosky:1960sp,Anastasiou:2002yz}, we express the on-shell delta functions as the cut propagators
\begin{align}
\delta(p^2)=\frac{1}{2\pi \img}\left(\frac{1}{p^2-\img0}-\frac{1}{p^2+\img0}\right)
\end{align}
and the measurement function as 
\begin{align}
\delta\Bigl(z-\frac{1-\cos\chi_{ij}}{2}\Bigr)&=\frac{p_i\cdot p_j}{z}\,\delta\bigl(2z(p_i\cdot Q)(p_j\cdot Q)-p_i\cdot p_j\bigr)\\
&= \frac{1}{2\pi \img} \frac{(p_i\cdot p_j)}{z}\biggl(\frac{1}{\left(2z(p_i\!\cdot\! Q)(p_j\!\cdot\! Q)\!-\!p_i\!\cdot\! p_j\right)\!-\!\img 0}-\frac{1}{\left(2z(p_i\!\cdot\! Q)(p_j\!\cdot\! Q)\!-\!p_i\!\cdot\! p_j\right)\!+\!\img 0}\biggr),
\nn \end{align}
where we set the center-of-mass energy $Q=(1,0,0,0)$ for simplicity (the dependence on $Q$ can be restored by dimensional analysis). The phase-space integrals can then be reduced to master integrals (MIs) using techniques from the study of multi-loop integrals. In particular, integration by parts and Lorentz invariance identities were generated with \texttt{LiteRed}~\cite{Lee:2012cn,Lee:2013mka} and the reduction to master integrals was performed using \texttt{FIRE6}~\cite{Smirnov:2019qkx}. The MIs are the same as that for the standard EEC and can be evaluated by the method of differential equations (DEs). The canonical forms of the DE systems are obtained by \texttt{CANONICA} \cite{Meyer:2017joq}. The solutions of the DEs are written in terms of harmonic polylogarithms, which can then be simplified to classical polylogarithms using \texttt{HPL}~\cite{Maitre:2005uu}. The calculation of $\Sigma_{c^p}$ is equivalent to the calculation of cut bubble integrals, and the master integrals can be found in refs.~\cite{Gehrmann-DeRidder:2003pne,Magerya:2019cvz}.

\subsubsection{Using Splitting Functions}\label{sec:calc_split}

While the calculation of the track function RG from the energy correlators provides a robust check on the universality of the track functions, it becomes computationally expensive at higher moments. Indeed, the main advantage of that approach, is that one also gets the full EEC distribution on tracks, which is itself a physically interesting observable. However, if one just wants the renormalization of the track functions, which is purely collinear in nature, it is easier to directly take advantage of collinear factorization, and obtain the RG from the splitting functions. Here we give a general description of this approach, with more details for the case of pure Yang-Mills given in app~\ref{app:yang_mills}. Although we focus in this paper on deriving moments, this approach has the added advantage that it can be generalized to allow a derivation of the full RG of the track functions in $x$-space.

To obtain a non-scaleless integral in the collinear limit, one must consider the measurement of some additional observable. We consider the measurement of the jet mass of \emph{all} particles and the energy fraction on charged particles, encoded in the jet function $J_a(s,x)$. The measurement of the jet mass renders the integrals non-scaleless, but importantly, the renormalization of $J_a(s,x)$ is identical to the standard $J_a(s)$ (see e.g.~\cite{Ritzmann:2014mka}). After performing this renormalization, as well as the standard renormalization of the strong coupling constant, the remaining poles determine the renormalization of the track functions.  
Unlike the pure gluon case considered in app.~\ref{app:yang_mills}, where all terms in the NLO evolution can be related to those involving three track functions, in the multi-flavor case, one must also consider terms involving two track functions. Therefore one must properly incorporate both the $1\to 3$ triple collinear splitting functions \cite{Campbell:1997hg,Catani:1998nv}, as well as the NLO corrections to the $1\to 2$ splitting functions \cite{Bern:1998sc,Bern:1999ry,Sborlini:2013jba}. 

We will now provide a bit more detail for each of these steps, starting with the calculation of the jet function $J_a(s,x)$: 
\begin{align}\label{eq:baretrackjet}
J_{a,\text{bare}}(s,x)=\sum_N \sum_{\{a_f\}}\int \df \Phi^c_N\, \delta(s-s')\, \sigma_{a \to \{a_f\}} ^c(\{z_f\},\{s_{f\!f'}\})\int \biggl[ \prod_{i=1}^N \df x_i T^{(0)}_{a_i}(x_i)\biggr]\,\delta \Bigl( x-\sum_{i=1}^N x_i z_i \Bigr)\,.
\end{align}
Here $\Phi^c_N$ is the $N$-particle collinear phase space with total invariant mass $s'$ and $\sigma_{a \to \{a_f\}}^c$ is the squared collinear matrix element for $a \to a_1 a_2 \cdots a_N$. At LO, $J_{\text{bare},f}^{(0)}(s,x)=\delta(s)T_f^{(0)}(x)$. The NLO calculation of the jet function gives rise to the LO RG evolution of the track functions. To derive the NLO RG for the track functions, we must consider the NNLO calculation of $J_{\text{bare}}(s,x)$.

At NNLO, we have both the NLO corrections to the two-particle final state (real-virtual corrections) and the three-particle final state (real-real corrections). Explicitly, 
\begin{align} \label{eq:J_NLO}
J_{a,\text{bare}}(s,x)\Big |_{a_s^2}\!&=\!\sum_{b,c} \int \!\df\Phi^c_2\, \delta(s\!-\!s') \sigma_{a\to bc}^c(z_b,z_c,s'\!=\!s_{bc})\int\! \df x_1 \df x_2 T^{(0)}_b(x_1) T^{(0)}_c(x_2) \delta(x\!-\!x_1 z_1\!-\!x_2 z_2)  
\nn \\ & \quad 
+ \sum_{b,c,d} \int \df \Phi^c_3 \delta(s-s') \sigma_{a\to bcd}^c(\{z_f\},\{s_{f\!f'}\})\int \df x_1 \df x_2 \df x_3 T^{(0)}_b(x_1) T^{(0)}_c(x_2) T^{(0)}_d(x_3) 
\nn \\ & \quad \times
\delta(x-x_1 z_1-x_2 z_2-x_3 z_3)\,,
\end{align}
where $\sigma_{a\to bc}^c$ and $\sigma_{a\to bcd}^c$ are the NLO $1\to2$ splitting and LO $1\to 3$ splitting functions respectively.

Taking moments of this equation
\begin{align}
J_a(s,n) \equiv \int\! \df x\, x^n J_a(s,x)\,,    
\end{align}
and using the sum rule for the track functions, one finds that $J_a(s,n)$ is expressed in terms of integrals of the $1\to 2$ and $1\to 3$ splitting functions weighted by a polynomial of weight $n$, as is done explicitly in app.~\ref{app:yang_mills} for the pure gluon case. These integrals can be performed explicitly using the approach of \cite{Kosower:2003np} (many integrals relevant for the quark case can be found in \cite{Ritzmann:2014mka}).

For each value of $n$, the renormalization of $J_a(s,n)$ in the variable $s$ is the same as for $J_a(s)$. Renormalizing the coupling using
\begin{align}
    Z_\alpha=1-\frac{\alpha_s}{4\pi}\frac{\beta_0}{\epsilon}+\mathcal{O}(\alpha_s^2)\,,
\end{align}
and expanding the bare jet function and the renormalization factor in terms of the \emph{renormalized} coupling, $J_{a,\text{bare}}=\sum_{L=0}^\infty a_s^L(\mu) J^{(L)}_{a,\text{bare}}$ and $Z_{J_a}=\sum_{L=0}^\infty a_s^L(\mu) Z_{J_a}^{(L)}$,  the two loop renormalization for the jet function is then 
\begin{align}\label{eq:jetRG_2loop}
J_a^{(2)}(s,n,\mu)= Z_{J_a}^{(2)}\otimes J^{(0)}_{a,\text{bare}}+Z_{J_a}^{(1)}\otimes J^{(1)}_{a,\text{bare}} 
+Z_{J_a}^{(0)}\otimes J^{(2)}_{a,\text{bare}}
\,.
\end{align}
The explicit form of the renormalization factors can be found in \cite{Ritzmann:2014mka} (for $a=q$) and \cite{Becher:2010pd} (for $a=g$) up to order $a_s^2$. After performing this renormalization in $s$, the RG for the track functions can be directly read off, as for the EEC based calculation in \Sec{sec:calc_correlator}. Explicitly, rewriting the tree-level track functions in \eqref{eq:J_NLO} in terms of the renormalized track functions, using $\vec {\bf T}_n^{(0)} = \vec {\bf T}_{n,{\rm bare}}$ and \eqref{eq:poles}, the UV poles from the renormalization in \eqref{eq:poles} should cancel against the IR poles from the direct integration in \eqref{eq:J_NLO}. This should be compared with the approach in app.~\ref{app:yang_mills}, which starts from the matching of the jet function onto renormalized track functions, where the matching coefficient is finite and the IR poles are contained in the track functions. Here, instead by expressing $T^{(0)}$ in terms of renormalized track functions, one automatically gets something of the form of a matching relation and the resulting coefficient is therefore the finite matching coefficient. Compared to the full EEC calculation, the integrals over the splitting functions are much easier (and mostly known). However, the fact that identical results are obtained from both approaches provides a strong check on our results.

\subsection{Results}\label{sec:res_explicit}

In this section we present results for the first six moments of the track functions. The results for the first three moments were presented in \cite{Li:2021zcf} and those for the fourth through sixth moments are new. These results are provided in electronic format accompanying this paper. We write the evolution equations for the central moments, whose definition can be found in \eqref{eq:centralmoment}, in terms of a perturbative expansion
\begin{align}
    \frac{\df}{\df\ln\mu^2}\sigma_a(N)
    &=
    \sum_{L=0}^\infty
    a_s^{L+1} D^{(L)}_{\sigma_a(N)} \ .
\end{align}

At a given order in perturbation theory there are constraints to which combinations of track functions can appear in the evolution equations. These constraints arise from the fact that in the evolution equation of $T_a$, a term involving the combination $T_b T_c$ originates from a $a\rightarrow bc X$ splitting contribution. The constraints from the possible splittings at a given order in perturbation theory results in linear dependencies between different terms in the evolution of central moments. This motivates the use of a minimal basis, in which both the constraints from possible splittings as well as shift invariance is clear throughout. For the evolution of the gluon central moments such a basis is provided by the following shift invariant quantity
\begin{align}
    \Delta_a (N)
    &=
    \int \df x \ \big[x - T_g(1)\big]^N T_a(x)
    =
    \sum_{k=0}^N \binom{N}{k} (-1)^k T_g^k (1) T_a(N-k) \ ,
\end{align}
while for the evolution of the quark central moments instead
\begin{align}
    \tau_a (N)
    &=
    \int \df x \ \big[x - T_q(1)\big]^N T_a(x)
    =
    \sum_{k=0}^N \binom{N}{k} (-1)^k T_q^k (1) T_a(N-k)
\end{align}
is used. The $\Delta_a$  introduced in sec.~\ref{sec:constr_multi} is equal to $\Delta_a(1)$, and we will abbreviate $\tau_a(1)=\tau_a$. Note that as a consequence of this notation, $\tau_g = -\Delta_q$.

The evolution of $\Delta_q$ is fixed to all loop orders in terms of the DGLAP anomalous dimension
\begin{align}
    D^{(n)}_{\Delta_q}
    &=
    -\big[\gamma_{gg}^{(n)}(2) + \gamma_{qq}^{(n)}(2)\big]\,{\color{blue} \Delta_q}\,.
\end{align}
The leading order evolution equations for gluons are given by
\begin{align} \label{eq:final_gluon_shift}
    D^{(0)}_{\sigma_g(2)} 
    & = 
    -\gamma^{(0)}_{gg}(3){\color{blue}\sigma_g(2)}
    +\sum_i\bigg\{
    -\gamma^{(0)}_{qg}(3){\color{blue}\left(\Delta_{q_i}(2)+\Delta_{\bar{q}_i}(2)\right)}
    +\frac{2}{5}T_F\, {\color{blue}\Delta_{q_i}\Delta_{\bar{q}_i}} 
    \bigg\} , 
    \nn \\
    D^{(0)}_{\sigma_g(3)} 
    &=  
    -\gamma^{(0)}_{gg}(4){\color{blue}\sigma_g(3)}
    +\sum_i\bigg\{
    -\gamma^{(0)}_{qg}(4){\color{blue}\Delta_{q_i}(3)} -2T_F\,{\color{blue}\sigma_g(2)\Delta_{q_i}} 
    +\frac{3}{10}T_F\,{\color{blue}\Delta_{q_i}(2)\Delta_{\bar{q}_i}}
    +(q_i\leftrightarrow\bar{q}_i)
    \bigg\} ,
    \nn\\
    D^{(0)}_{\sigma_g(4)} 
    &=  
    -\gamma^{(0)}_{gg}(5){\color{blue}\sigma_g(4)}
    +\frac{9}{7} C_A \,{\color{blue}\sigma_g^2(2)}
    +\sum_i\bigg\{
    -\gamma^{(0)}_{qg}(5){\color{blue}\Delta_{q_i}(4)} 
    +\frac{26}{105} T_F \,{\color{blue}\Delta_{q_i}(3) \Delta_{\bar{q}_i}}
    \nn\\ 
    &\quad  
    +\frac{4}{35} T_F \,{\color{blue}\Delta_{q_i}(2) \Delta_{\bar{q}_i}(2)}
    -\frac{8}{3} T_F \,{\color{blue}\sigma_g(3) \Delta_{q_i}}
    +(q_i\leftrightarrow\bar{q}_i)
    \bigg\} ,
    \nn\\
    D^{(0)}_{\sigma_g(5)} 
    &=  
    -\gamma^{(0)}_{gg}(6){\color{blue}\sigma_g(5)}
    +\frac{15}{7} C_A \,{\color{blue}\sigma_g(3) \sigma_g(2)}
    +\sum_i\bigg\{
    -\gamma^{(0)}_{qg}(6){\color{blue}\Delta_{q_i}(5)} 
    +\frac{3}{14} T_F \,{\color{blue}\Delta_{q_i}(4) \Delta_{\bar{q}_i}}
    \nn\\ 
    &\quad  
    +\frac{4}{21} T_F \,{\color{blue}\Delta_{q_i}(3) \Delta_{\bar{q}_i}(2)}
    -\frac{10}{3} T_F \,{\color{blue}\sigma_g(4) \Delta_{q_i}}
    +(q_i\leftrightarrow\bar{q}_i)
    \bigg\} ,
    \nn\\
    D^{(0)}_{\sigma_g(6)} 
    &=  
    -\gamma^{(0)}_{gg}(7){\color{blue}\sigma_g(6)}
    +\frac{83}{42} C_A \,{\color{blue}\sigma_g(4) \sigma_g(2)}
    +\frac{52}{63} C_A \,{\color{blue}\sigma^2_g(3)}
    \nn\\
    &\quad
    +\sum_i\bigg\{
    -\gamma^{(0)}_{qg}(7){\color{blue}\Delta_{q_i}(6)} 
    +\frac{4}{21} T_F \,{\color{blue}\Delta_{q_i}(5) \Delta_{\bar{q}_i}}
    +\frac{1}{6} T_F \,{\color{blue}\Delta_{q_i}(4) \Delta_{\bar{q}_i}(2)}
    \nn\\
    &\qquad
    +\frac{5}{63} T_F \,{\color{blue}\Delta_{q_i}(3) \Delta_{\bar{q}_i}(3)}
    -4T_F \,{\color{blue}\sigma_g(5) \Delta_{q_i}}
    +(q_i\leftrightarrow\bar{q}_i)
    \bigg\} ,
\end{align}
and for quarks they are given by
\begin{align}
    D^{(0)}_{\sigma_q(2)}
    &=
    -\gamma^{(0)}_{qq}(3) {\color{blue} \sigma_q(2)}
    -\gamma^{(0)}_{gq}(3) {\color{blue} \tau_g(2)} \ ,
    \nn\\
    D^{(0)}_{\sigma_q(3)}
    &=
    -\gamma^{(0)}_{qq}(4) {\color{blue} \sigma_q(3)}
    -\gamma^{(0)}_{gq}(4) {\color{blue} \tau_g(3)}
    -\frac{24}{5} C_F \,{\color{blue} \sigma_q(2) \tau_g} \ ,
    \nn\\
    D^{(0)}_{\sigma_q(4)}
    &=
    -\gamma^{(0)}_{qq}(5) {\color{blue} \sigma_q(4)}
    -\gamma^{(0)}_{gq}(5) {\color{blue} \tau_g(4)}
    -\frac{22}{3} C_F \,{\color{blue} \sigma_q(3) \tau_g}
    +\frac{7}{5} C_F \,{\color{blue} \sigma_q(2) \tau_g(2)} \ ,
    \nn\\
    D^{(0)}_{\sigma_q(5)}
    &=
    -\gamma^{(0)}_{qq}(6) {\color{blue} \sigma_q(5)}
    -\gamma^{(0)}_{gq}(6) {\color{blue} \tau_g(5)}
    -\frac{208}{21} C_F \,{\color{blue} \sigma_q(4) \tau_g}
    +\frac{31}{21} C_F \,{\color{blue} \sigma_q(3) \tau_g(2)}
    +\frac{6}{7} C_F \,{\color{blue} \sigma_q(2) \tau_g(3)} \ ,
    \nn \\
    D^{(0)}_{\sigma_q(6)}
    &=
    -\gamma^{(0)}_{qq}(7) {\color{blue} \sigma_q(6)}
    -\gamma^{(0)}_{gq}(7) {\color{blue} \tau_g(6)}
    -\frac{25}{2} C_F \,{\color{blue} \sigma_q(5) \tau_g}
    \nn \\
    &\quad
    +\frac{43}{28} C_F \,{\color{blue} \sigma_q(4) \tau_g(2)}
    +\frac{19}{21} C_F \,{\color{blue} \sigma_q(3) \tau_g(3)}
    +\frac{17}{28} C_F \,{\color{blue} \sigma_q(2) \tau_g(4)} \ .
\end{align}
At NLO the evolution equations for the first six moments of the gluon track functions are
\begin{align}
D^{(1)}_{\sigma_g(2)}
    &=
    -\gamma_{gg}^{(1)}(3) {\color{blue} \sigma_g(2)}
    +\sum_i\bigg\{
    -\gamma_{qg}^{(1)}(3) 
    \,{\color{blue} \left(\Delta_{q_i}(2) + \Delta_{\bar{q}_i}(2)\right)}
    \\
    &\quad\qquad
    +T_F\bigg[\bigg(\frac{12413}{1350}-\frac{5 2 \pi^2}{45}\bigg) C_A 
    +\frac{1528}{225} C_F -\frac{16}{25} n_f T_F\bigg]
    \,{\color{blue} \Delta_{q_i} \Delta_{\bar{q}_i}}
    \bigg\},
    \nn\\
    D^{(1)}_{\sigma_g(3)}
    &=
    -\gamma_{gg}^{(1)}(4) {\color{blue} \sigma_g(3)}
    +\sum_i\bigg\{
    \!-\!\gamma_{qg}^{(1)}(4) {\color{blue} \Delta_{q_i}(3)}
    +\!T_F\bigg[\bigg(\!-\!\frac{638}{45}\!+\!\frac{8\pi^2}{3}\bigg) C_A\!-\!\frac{3803}{250} C_F\bigg]
    {\color{blue} \sigma_g(2) \Delta_{q_i}}
    \nn\\
    &\qquad
    +T_F\bigg[\bigg(\frac{5321}{3000} -\frac{2\pi^2}{5}\bigg) C_A
    +\frac{1523}{240} C_F -\frac{12}{25} n_F T_F\bigg]
    \,{\color{blue} \Delta_{q_i}(2) \Delta_{\bar{q}_i}}
    +(q_i \leftrightarrow \bar{q}_i)
    \bigg\},
    \nn\\
    D^{(1)}_{\sigma_g(4)}
    &=
    -\gamma_{gg}^{(1)}(5) {\color{blue} \sigma_g(4)}
    +\bigg[\left(\frac{20709772}{55125}-\frac{1584 \pi^2}{35} + 72\zeta_3\right)C_A^2
    -\frac{4}{15}C_A n_f T_F\bigg]
    \,{\color{blue} \sigma_g(2)^2}
    \nn\\
    &\quad
    +\sum_i \bigg\{
    -\gamma_{qg}^{(1)}(5) \,{\color{blue} \Delta_{q_i}(4)}
    +T_F\bigg[\left(-\frac{66482}{3675} + \frac{32\pi^2}{9}\right) C_A 
    -\frac{1291307}{66150} C_F\bigg] 
    \,{\color{blue} \sigma_g(3) \Delta_{q_i}}
    \nn\\
    &\qquad
    +T_F\bigg[\left(-\frac{51721}{2625} + \frac{28\pi^2}{15}\right) C_A
    -\frac{5177 }{7875}C_F\bigg] 
    \,{\color{blue} \sigma_g(2) \Delta_{q_i}(2)}
    \nn\\
    &\qquad
    +T_F\bigg[\left(\frac{1018886}{55125}-\frac{28\pi^2}{15}\right) C_A
    -\frac{11889}{24500} C_F\bigg]
    \,{\color{blue} \sigma_g(2) \Delta_{q_i} \Delta_{\bar{q}_i}}
    \nn\\
    &\qquad
    +T_F\bigg[\left(\frac{22403}{2450}-\frac{8\pi^2}{7}\right) C_A 
    +\frac{3794489}{661500} C_F -\frac{1136}{3675} n_f T_F\bigg]
    \,{\color{blue} \Delta_{q_i}(3) \Delta_{\bar{q}_i}}
    \nn\\
    &\qquad
    +T_F\bigg[\left(\!-\!\frac{68429}{12250}\!+\!\frac{16\pi^2}{35}\right) C_A 
    \!+\!\frac{35003}{11025} C_F \!-\!\frac{304}{1225} n_f T_F 
    \bigg]
    \,{\color{blue} \Delta_{q_i}(2) \Delta_{\bar{q}_i}(2)}
    +(q_i \leftrightarrow \bar{q}_i)
    \bigg\},
    \nn\\
    D^{(1)}_{\sigma_g(5)}
    &=
    -\gamma_{gg}^{(1)}(6) {\color{blue} \sigma_g(5)}
    +\bigg[\bigg(\frac{239432987}{617400}\!-\!\frac{2896\pi^2}{63}\!+\!80\zeta_3\bigg) 
    C_A^2 -\frac{4}{9} C_A n_f T_F\bigg]
    \,{\color{blue} \sigma_g(3) \sigma_g(2)}
    \nn\\
    &\quad
    +\sum_i\bigg\{
    -\gamma_{qg}^{(1)}(6) {\color{blue} \Delta_{q_i}(5)}
    +T_F\bigg[\bigg(-\frac{579361}{26460}+\frac{40 \pi^2}{9}\bigg) C_A  
    -\frac{11205259}{463050} C_F\bigg] 
    \,{\color{blue} \sigma_g(4) \Delta_{q_i}}
    \nn\\
    &\qquad
    +T_F\bigg[\bigg(-\frac{202039}{6300}+\frac{28 \pi^2}{9}\bigg) C_A
    -\frac{68329}{308700} C_F\bigg] 
    \,{\color{blue} \sigma_g(3) \Delta_{q_i}(2)}
    \nn\\
    &\qquad
    +T_F\bigg[\bigg(\frac{579007}{18900}-\frac{28 \pi^2}{9}\bigg)C_A 
    -\frac{449}{2450}C_F\bigg] 
    \,{\color{blue} \sigma_g(3) \Delta_{q_i} \Delta_{\bar{q}_i}}
    \nn\\
    &\qquad
    +T_F\bigg[-\frac{45197}{52920} C_A -\frac{41605}{74088} C_F\bigg] 
    \,{\color{blue} \sigma_g(2) \Delta_{q_i}(3)}
    \nn\\
    &\qquad
    +T_F\bigg[\frac{4499}{17640} C_A - \frac{1839}{1960} C_F\bigg] 
    \,{\color{blue} \sigma_g(2) \Delta_{q_i}(2) \Delta_{\bar{q}_i}}
    \nn\\
    &\qquad
    +T_F\bigg[\bigg(\frac{114511}{148176}\!-\!\frac{2 \pi^2}{7}\bigg) C_A
    \!+\!\frac{152459}{29400}C_F \!-\!\frac{44}{245} n_f T_F\bigg]
    \,{\color{blue} \Delta_{q_i}(4) \Delta_{\bar{q}_i}}
    \nn\\
    &\qquad
    +T_F\bigg[\bigg(\frac{34183}{92610}\!-\!\frac{16 \pi^2}{63}\bigg)C_A
    \!+\!\frac{198559}{33075} C_F \!-\!\frac{304}{735} n_f T_F\bigg] 
    \,{\color{blue} \Delta_{q_i}(3) \Delta_{\bar{q}_i}(2)}
    +(q_i \leftrightarrow \bar{q}_i)
    \bigg\},
    \nn \\
    D^{(1)}_{\sigma_g(6)}
    &=
    -\gamma_{gg}^{(1)}(7) {\color{blue} \sigma_g(6)}
    +C_A^2 \bigg[\frac{299405789}{137200}-\frac{1856 \pi^2}{7}+360\zeta_3\bigg]
    \,{\color{blue} \sigma^3_g(2)}
    \nn \\
    &\quad
    +\bigg[
    \left(-\frac{3348739}{6075}+\frac{1810 \pi^2}{27}-80\zeta_3\right)C_A^2 -\frac{4}{21}C_A n_f T_F\bigg] 
    \,{\color{blue} \sigma_g(3)^2}
    \nn\\
    &\quad
    +\bigg[\left(\frac{47613060961}{22226400}-\frac{2321
    \pi^2}{9}+360\zeta_3\right)C_A^2 -\frac{8}{21} C_A n_f T_F\bigg]
    \,{\color{blue} \sigma_g(4) \sigma_g(2)}
    \nn\\
    &\quad
    +\sum_i\bigg\{
    \!-\!\gamma_{qg}^{(1)}(7) \Delta_{q_i}(6)
    +T_F\bigg[\left(\!-\!\frac{10192933}{396900}\!+\!\frac{16 \pi^2}{3}\right)C_A 
    \!-\!\frac{91953847}{3175200} C_F\bigg] 
    \,{\color{blue} \sigma_g(5) \Delta_{q_i}}
    \nn\\
    &\qquad
    +T_F\bigg[\left(-\frac{75307691}{1587600}+\frac{14 \pi^2}{3}\right)C_A 
    -\frac{4613227}{44452800} C_F\bigg]
    \,{\color{blue} \sigma_g(4) \Delta_{q_i}(2)}
    \nn\\
    &\qquad
    +T_F\bigg[\left(\frac{24317347}{529200}-\frac{14 \pi^2}{3}\right)C_A
    -\frac{17153}{185220} C_F\bigg]
    \,{\color{blue} \sigma_g(4) \Delta_{q_i} \Delta_{\bar{q}_i}}
    \nn\\
    &\qquad
    +T_F\bigg[-\frac{2128943}{2381400}C_A-\frac{1218841}{6667920} C_F\bigg]
    \,{\color{blue} \sigma_g(3) \Delta_{q_i}(3)}
    \nn\\
    &\qquad
    +T_F\bigg[\frac{947}{264600}C_A-\frac{84409}{231525} C_F\bigg] 
    \,{\color{blue} \sigma_g(3) \Delta_{q_i}(2) \Delta_{\bar{q}_i}}
    \nn\\
    &\qquad
    +T_F\bigg[\left(-\frac{42709397}{1389150}+\frac{64 \pi^2}{21}\right)C_A
    -\frac{4334179}{8890560} C_F\bigg] 
    \,{\color{blue} \sigma_g(2) \Delta_{q_i}(4)}
    \nn\\
    &\qquad
    +T_F\bigg[\left(\frac{669778843}{5556600}-\frac{256 \pi^2}{21}\right)C_A 
    -\frac{48397}{55566} C_F\bigg] 
    \,{\color{blue} \sigma_g(2) \Delta_{q_i}(3) \Delta_{\bar{q}_i}}
    \nn\\
    &\qquad
    +T_F\bigg[\left(-\frac{16674109}{185220}+\frac{64 \pi^2}{7}\right)C_A  
    -\frac{36343}{75600} C_F\bigg]
    \,{\color{blue} \sigma_g(2) \Delta_{q_i}(2) \Delta_{\bar{q}_i}(2)}
    \nn\\
    &\qquad
    +T_F\bigg[\left(\frac{53650579}{5556600}\!-\!\frac{74 \pi^2}{63}\right)C_A
    \!+\!\frac{6547967}{1389150} C_F
    \!-\!\frac{1684}{19845} n_f T_F\bigg]
    \,{\color{blue} \Delta_{q_i}(5) \Delta_{\bar{q}_i}}
    \nn\\
    &\qquad
    +T_F\bigg[\left(\!-\!\frac{502728871}{22226400}\!+\!\frac{131 \pi^2}{63}\right)C_A
    \!+\!\frac{364099}{64800} C_F
    \!-\!\frac{926}{2835} n_f T_F\bigg]
    \,{\color{blue} \Delta_{q_i}(4) \Delta_{\bar{q}_i}(2)}
    \nn\\
    &\qquad
    +T_F\bigg[\left(\frac{100850479}{6667920}\!-\!\frac{310\pi^2}{189}\right)C_A
    \!+\!\frac{14171}{4860} C_F 
    \!-\!\frac{2332}{11907} n_f T_F\bigg]
    \,{\color{blue} \Delta_{q_i}(3) \Delta_{\bar{q}_i}(3)}
    +(q_i \leftrightarrow \bar{q}_i)
    \bigg\}, \nn
\end{align}
and for the quark track functions
\begin{align}
D^{(1)}_{\sigma_q(2)}
    &=
    -\gamma_{gq}^{(1)}(3) {\color{blue} \tau_g(2)}
    -\gamma_{qq}^{(1)}(3) {\color{blue} \sigma_q(2)}
    -\gamma_{\bar{q}q}^{(1)}(3) \,{\color{blue} \tau_{\bar{q}}(2)}
     + \left[\left(\frac{1399}{5400}-\frac{7 \pi ^2}{9}\right)
   C_A C_F-\frac{67 }{18}C_F^2\right] {\color{blue} \tau_g^2 }
    \nn\\
    &\quad
    +\sum_i \bigg\{
    -\gamma_{Qq}^{(1)}(3) {\color{blue} \left(\tau_{Q_i}(2)+\tau_{\bar{Q}_i}(2)\right)}
    -\frac{17}{100} C_F T_F \,{\color{blue} \tau_{Q_i} \tau_{\bar{Q}_i}}
    \bigg\} ,
    \nn\\
    D^{(1)}_{\sigma_q(3)}
    &=
    -\gamma_{gq}^{(1)}(4) {\color{blue} \tau_g(3)}
    -\gamma_{qq}^{(1)}(4) {\color{blue} \sigma_q(3)}
    -\gamma_{\bar{q}q}^{(1)}(4) {\color{blue} \tau_{\bar{q}}(3)}
    \nn\\
    &\quad
    +\bigg[\bigg(\frac{1204633}{18000}-\frac{247\pi^2}{30} + 12\zeta_3\bigg)\frac{C_F}{N_c}
    +\frac{11503}{3000} C_F T_F\bigg]
    \,{\color{blue} \sigma_q(2) \tau_{\bar{q}}}
    \nn\\
    &\quad
    +\bigg[\bigg(\frac{32 \pi^2}{5}-\frac{50299}{2250}\bigg)C_F^2 
    -\frac{577}{20} C_A C_F\bigg] 
    \,{\color{blue} \sigma_q(2) \tau_{g}}
    \nn\\
    &\quad
    +\bigg[-\frac{249}{50} C_F^2 -\frac{3787}{750} C_A C_F\bigg]
    \,{\color{blue} \tau_g(2) \tau_g}
    \nn\\
    &\quad
    +C_F T_F\sum_i\bigg\{
    -\frac{11867}{27000}\,{\color{blue} \tau_{Q_i}(3)}
    +\frac{292}{75}\,{\color{blue} \sigma_q(2) \tau_{Q_i}}
    -\frac{59}{1000} \,{\color{blue} \tau_{Q_i}(2) \tau_{\bar{Q}_i}}
    +(Q_i \leftrightarrow \bar{Q}_i)
    \bigg\} ,
    \nn\\
    D^{(1)}_{\sigma_q(4)}
    &=
    -\gamma_{qq}^{(1)}(5) {\color{blue} \sigma_q(4)}
    -\gamma_{gq}^{(1)}(5) {\color{blue} \tau_g(4)}
    -\gamma_{\bar{q}q}^{(1)}(5) {\color{blue} \tau_{\bar{q}}(4)}
    \nn\\
    &\quad
    +\bigg[\left(\frac{88 \pi ^2}{9}-\frac{109699}{2700}\right)
    C_F^2-\frac{1061}{25} C_A C_F\bigg] \,{\color{blue} \sigma_q(3) \tau_g}
    \nn\\
    &\quad
    +\bigg[\left(\frac{151903}{9000}-\frac{28 \pi ^2}{15}\right) C_F^2
    +\left(\frac{14 \pi ^2}{3}-\frac{114827}{4500}\right) C_A C_F\bigg]
    \,{\color{blue} \sigma_q(2) \tau_g(2)}
    \nn\\
    &\quad
    +\bigg[-\frac{29}{45} C_F^2 
    +\left(\frac{141647}{4500}-\frac{14 \pi ^2}{3}\right) C_A C_F\bigg] 
    \,{\color{blue} \sigma_q(2) \tau^2_g}
    \nn\\
    &\quad
    +\bigg[-\frac{67}{18} C_F^2
    +\left(\frac{1795559}{165375}-\frac{64 \pi^2}{45}\right) C_A C_F\bigg] 
    \,{\color{blue} \tau_g(3) \tau_g}
    \nn\\
    &\quad
    +\bigg[-\frac{3143}{1800} C_F^2
    +\left(\frac{16 \pi ^2}{15}-\frac{593207}{55125}\right) C_A C_F\bigg] 
    \,{\color{blue} \tau_g(2)^2}
    \nn\\
    &\quad
    +\bigg[\left(-\frac{203009}{1800}+\frac{43\pi^2}{3} -24\zeta_3\right)\frac{C_F}{N_c}
    -\frac{5323}{4500} C_F T_F\bigg]
    \,{\color{blue} \sigma_q(2)^2}
    \nn\\
    &\quad
    +\bigg[\left(\frac{676639}{6750}-\frac{544\pi^2}{45}+16\zeta_3\right)\frac{C_F}{N_c}
    +\frac{998092}{165375} C_F T_F\bigg]
    \,{\color{blue} \sigma_q(3) \tau_{\bar{q}}}
    \nn\\
    &\quad
    +\bigg[\left(-\frac{2364943}{9000}+\frac{487\pi^2}{15} -48\zeta_3\right)\frac{C_F}{N_c}
    -\frac{258203}{220500} C_F T_F\bigg]
    \,{\color{blue} \sigma_q(2) \tau_{\bar{q}}(2)}
    \nn\\
    &\quad
    +C_F T_F\sum_i
    \bigg\{-\frac{46516}{165375} \,{\color{blue} \tau_{Q_i}(4)}
    -\frac{7633}{165375} \,{\color{blue} \tau_{Q_i}(3) \tau_{\bar{Q}_i}}
    +\frac{328}{55125} \,{\color{blue} \tau_{Q_i}(2) \tau_{\bar{Q}_i}(2) }
    \nn\\
    &\qquad+\frac{821}{135} \,{\color{blue} \sigma_q(3) \tau_{Q_i}}
    -\frac{5323}{4500} \,{\color{blue} \sigma_q(2) \tau_{Q_i}(2)}
    -\frac{229}{1500} \,{\color{blue} \sigma_q(2) \tau_{Q_i} \tau_{\bar{Q}_i}}
    +(Q_i \leftrightarrow \bar{Q}_i)\bigg\} ,
    \nn \\
    D^{(1)}_{\sigma_q(5)}
    &=
    -\gamma_{qq}^{(1)}(6) {\color{blue} \sigma_q(5)}
    -\gamma_{gq}^{(1)}(6) {\color{blue} \tau_g(5)}
    -\gamma_{gq}^{(1)}(6) {\color{blue} \tau_{\bar{q}}(5)}
    \nn\\
    &\quad
    +\bigg[\left(-\frac{13864028}{231525}+\frac{832 \pi
    ^2}{63}\right)C_F^2 -\frac{370753}{6615} C_A C_F\bigg] 
    \,{\color{blue} \sigma_q(4) \tau_g(1)}
    \nn\\
    &\quad
    +\bigg[\left(\frac{11490901}{617400}-\frac{124 \pi ^2}{63}\right)
    C_F^2+\left(-\frac{2406319}{44100}+\frac{70 \pi ^2}{9}\right) C_A C_F\bigg] 
    \,{\color{blue} \sigma_q(3) \tau_g(2)}
    \nn\\
    &\quad
    +\bigg[\left(\frac{3038951}{308700}-\frac{8 \pi ^2}{7}\right)
    C_F^2+\frac{1064053}{66150} C_A C_F\bigg] 
    \,{\color{blue} \sigma_q(2) \tau_g(3)}
    \nn\\
    &\quad
    +\bigg[-\frac{32762}{11025} C_F^2-\frac{552788}{231525} C_A C_F\bigg] 
    \,{\color{blue} \tau_g(4) \tau_g(1)}
    \nn\\
    &\quad
    +\bigg[-\frac{29527}{11025} C_F^2 + \frac{1606}{46305} C_A C_F\bigg] 
    \,{\color{blue} \tau_g(3) \tau_g(2)}
    \nn\\
    &\quad
    +\bigg[-\frac{19}{45} C_F^2 + \left(\frac{8012047}{132300}-\frac{70 \pi
    ^2}{9}\right)C_A C_F \bigg] 
    \,{\color{blue} \sigma_q(3) \tau_g(1)^2}
    \nn\\
    &\quad
    +\bigg[-\frac{44}{45} C_F^2-\frac{133918}{11025} C_A C_F\bigg] 
    \,{\color{blue} \sigma_q(2) \tau_g(2) \tau_g(1)}
    \nn\\
    &\quad
    +\bigg[\left(-\frac{54341821}{132300}+\frac{154 \pi ^2}{3}-80\zeta_3\right)
    \frac{C_F}{N_c}-\frac{63706}{33075} C_F T_F\bigg] 
    \,{\color{blue} \sigma_q(3) \sigma_q(2)}
    \nn\\
    &\quad
    +\bigg[\left(\frac{58946437}{88200}-\frac{247 \pi ^2}{3}+120\zeta_3\right)
    \frac{C_F}{N_c}-\frac{667}{4900} C_F T_F\bigg] 
    \,{\color{blue} \sigma_q(2)^2 \tau_{\bar{q}}(1)}
    \nn\\
    &\quad
    +\bigg[\left(\frac{253049689}{1852200}-\frac{1024 \pi ^2}{63}+20\zeta_3\right)
    \frac{C_F}{N_c}+\frac{3832909}{463050} C_F T_F\bigg] 
    \,{\color{blue} \sigma_q(4) \tau_{\bar{q}}(1)}
    \nn\\
    &\quad
    +\bigg[\left(-\frac{852928693}{1852200}+\frac{3553 \pi ^2}{63}-80\zeta_3\right)
    \frac{C_F}{N_c}-\frac{232447}{185220} C_F T_F\bigg] 
    \,{\color{blue} \sigma_q(3) \tau_{\bar{q}}(2)}
    \nn\\
    &\quad
    +\bigg[\left(\frac{305938939}{617400}-\frac{1259 \pi^2}{21}+80\zeta_3 
    \right)\frac{C_F}{N_c}-\frac{66229}{102900} C_F T_F\bigg] 
    \,{\color{blue} \sigma_q(2) \tau_{\bar{q}}(3)}
    \nn\\
    &\quad
    +C_F T_F\sum_i \bigg\{
    -\frac{3649}{18522} \,{\color{blue} \tau_{Q_i}(5)}
    -\frac{6257}{154350} \,{\color{blue} \tau_{Q_i}(4) \tau_{\bar{Q}_i}(1)}
    +\frac{3184}{231525} \,{\color{blue} \tau_{Q_i}(3) \tau_{\bar{Q}_i}(2)}
    \nn\\
    &\qquad
    -\frac{86971}{132300} \,{\color{blue} \sigma_q(2) \tau_{Q_i}(3)}
    -\frac{667}{4900} \,{\color{blue} \sigma_q(2) \tau_{Q_i}(2) \tau_{\bar{Q}_i}(1)}
    -\frac{7219}{44100} \,{\color{blue} \sigma_q(3) \tau_{\bar{Q}_i}(1) \tau_{Q_i}(1)}
    \nn\\
    &\qquad
    -\frac{7993}{6300} \,{\color{blue} \sigma_q(3) \tau_{Q_i}(2)}
    +\frac{55024}{6615} \,{\color{blue} \sigma_q(4) \tau_{Q_i}(1)}
    +(Q_i \leftrightarrow \bar{Q}_i)
    \bigg\},
    \nn\\
    D^{(1)}_{\sigma_q(6)}
    &=
    -\gamma_{qq}^{(1)}(7) {\color{blue} \sigma_q(6)}
    -\gamma_{gq}^{(1)}(7) {\color{blue} \tau_g(6)}
    -\gamma_{\bar{q}q}^{(1)}(7) {\color{blue} \tau_{\bar{q}}(6)}
    \nn\\
    &\quad
    +\bigg[\left(-\frac{6245817}{78400}+\frac{50 \pi^2}{3}\right) C_F^2
    -\frac{4098089}{58800} C_A C_F\bigg] 
    \,{\color{blue} \sigma_q(5) \tau_g(1)}
    \nn\\
    &\quad
    +\bigg[\left(\frac{81534493}{7408800}-\frac{76 \pi^2}{63}\right)C_F^2
    +\frac{9352657}{529200} C_A C_F\bigg] 
    \,{\color{blue} \sigma_q(3) \tau_g(3)}
    \nn\\
    &\quad
    +\bigg[\left(\frac{195862451}{9878400}-\frac{43\pi^2}{21}\right) C_F^2
    +\left(-\frac{2309563}{25200}+\frac{35\pi^2}{3}\right) C_A C_F\bigg] 
    \,{\color{blue} \sigma_q(4) \tau_g(2)}
    \nn\\
    &\quad
    +\bigg[\left(\frac{22075117}{3292800}-\frac{17 \pi^2}{21}\right)C_F^2
    +\left(-\frac{3925723}{98784}+\frac{16 \pi^2}{3}\right) C_A C_F\bigg] 
    \,{\color{blue} \sigma_q(2) \tau_g(4)}
    \nn\\
    &\quad
    +\bigg[-\frac{1801}{5880} C_F^2
    +\left(\frac{4926391}{50400}-\frac{35 \pi^2}{3}\right) C_A C_F\bigg] 
    \,{\color{blue} \sigma_q(4) \tau_g(1)^2}
    \nn\\
    &\quad
    +\bigg[-\frac{641}{980} C_F^2-\frac{812183}{58800} C_A C_F\bigg] 
    \,{\color{blue} \sigma_q(3) \tau_g(2) \tau_g(1)}
    \nn\\
    &\quad
    +\bigg[
    -\frac{67507}{88200} C_F^2
    +\left(\frac{249991487}{1234800}-\frac{64 \pi^2}{3}\right) C_A C_F\bigg] 
    \,{\color{blue} \sigma_q(2) \tau_g(3) \tau_g(1)}
    \nn\\
    &\quad
    +\bigg[
    -\frac{15761}{39200} C_F^2
    +\left(-\frac{65733319}{411600}+16 \pi^2\right) C_A C_F\bigg] 
    \,{\color{blue} \sigma_q(2) \tau_g(2)^2}
    \nn\\
    &\quad
    +\bigg[
    -\frac{2183}{882} C_F^2
    +\left(\frac{129951169}{11113200}-\frac{29 \pi^2}{21}\right) C_A C_F\bigg] 
    \,{\color{blue} \tau_g(5) \tau_g(1)}
    \nn\\
    &\quad
    +\bigg[
    -\frac{1533479}{705600} C_F^2
    +\left(-\frac{60361165}{1778112}+\frac{145 \pi^2}{42}\right) C_A C_F\bigg] 
    \,{\color{blue} \tau_g(4) \tau_g(2)}
    \nn\\
    &\quad
    +\bigg[-\frac{11488}{11025} C_F^2
    +\left(\frac{8722057}{381024}-\frac{145 \pi^2}{63}\right) C_A C_F\bigg] 
    \,{\color{blue} \tau_g(3)^2}
    \nn\\
    &\quad
    +\bigg[\left(\frac{4333961129}{24696000}-\frac{1447 \pi^2}{70}+24\zeta_3\right)
    \frac{C_F}{N_c}+\frac{2344837}{222264} C_F T_F\bigg] 
    \,{\color{blue} \sigma_q(5) \tau_{\bar{q}}(1)}
    \nn\\
    &\quad
    +\bigg[\left(-\frac{66232429}{100800}+\frac{487 \pi^2}{6}-120\zeta_3\right)
    \frac{C_F}{N_c}-\frac{4355153}{2469600} C_F T_F\bigg] 
    \,{\color{blue} \sigma_q(4) \sigma_q(2)}
    \nn\\
    &\quad
    +\bigg[\left(-\frac{2369204633}{3292800}+\frac{7351 \pi^2}{84}-120\zeta_3\right)
    \frac{C_F}{N_c}-\frac{8430209}{6350400} C_F T_F\bigg] 
    \,{\color{blue} \sigma_q(4) \tau_{\bar{q}}(2)}
    \nn\\
    &\quad
    +\bigg[\left(-\frac{26205797}{58800}+\frac{494 \pi^2}{9}-80\zeta_3\right)
    \frac{C_F}{N_c}-\frac{41633}{58800} C_F T_F\bigg] 
    \,{\color{blue} \sigma_q(3)^2}
    \nn\\
    &\quad
    +\bigg[\left(\frac{7626957731}{7408800}-\frac{7799 \pi^2}{63}+160\zeta_3\right)
    \frac{C_F}{N_c}-\frac{11452643}{16669800} C_F T_F\bigg] 
    \,{\color{blue} \sigma_q(3) \tau_{\bar{q}}(3)}
    \nn\\
    &\quad
    +\bigg[\left(\frac{500596781}{176400}-346 \pi^2+480\zeta_3\right)\frac{C_F}{N_c}
    -\frac{3926}{15435} C_F T_F\bigg]
    \,{\color{blue} \sigma_q(3) \sigma_q(2) \tau_{\bar{q}}(1)}
    \nn\\
    &\quad
    +\bigg[\left(-\frac{8033396911}{9878400}+\frac{8149 \pi^2}{84}-120\zeta_3\right)
    \frac{C_F}{N_c}- \frac{18635861}{44452800} C_F T_F\bigg] 
    \,{\color{blue} \sigma_q(2) \tau_{\bar{q}}(4)}
    \nn\\
    &\quad
    +\bigg[\left(-\frac{264811033}{117600}+272 \pi^2-360\zeta_3\right) \frac{C_F}{N_c}
    -\frac{2761}{102900} C_F T_F\bigg] 
    \,{\color{blue} \sigma_q(2)^2 \tau_{\bar{q}}(2)}
    \nn\\
    &\quad
    +C_F T_F\sum_i\bigg\{
    -\frac{779767}{5334336} \,{\color{blue} \tau_{Q_i}(6)}
    -\frac{3949}{111132} \,{\color{blue} \tau_{Q_i}(5) \tau_{\bar{Q}_i}(1)}
    +\frac{10649}{1270080} \,{\color{blue} \tau_{Q_i}(4) \tau_{\bar{Q}_i}(2)}
    \nn\\
    &\qquad
    +\frac{28025}{2667168} \,{\color{blue} \tau_{Q_i}(3) \tau_{\bar{Q}_i}(3)}
    -\frac{33001}{77175} \,{\color{blue} \sigma_q(2) \tau_{Q_i}(4)}
    -\frac{135433}{1234800} \,{\color{blue} \sigma_q(2) \tau_{Q_i}(3) \tau_{\bar{Q}_i}(1)}
    \nn\\
    &\qquad
    -\frac{2761}{205800} \,{\color{blue} \sigma_q(2) \tau_{Q_i}(2) \tau_{\bar{Q}_i}(2)}
    -\frac{41633}{58800} \,{\color{blue} \sigma_q(3) \tau_{Q_i}(3)}
    -\frac{8507}{58800} \,{\color{blue} \sigma_q(3) \tau_{Q_i}(2) \tau_{\bar{Q}_i}(1)}
    \nn\\
    &\qquad
    -\frac{7481}{5600} \,{\color{blue} \sigma_q(4) \tau_{Q_i}(2)}
    -\frac{20369}{117600} \,{\color{blue} \sigma_q(4) \tau_{Q_i}(1) \tau_{\bar{Q}_i}(1)}
    +\frac{5335}{504} \,{\color{blue} \sigma_q(5) \tau_{Q_i}(1)}
    +(Q_i \leftrightarrow \bar{Q}_i)
    \bigg\} .
\end{align}
This evolution in moment space is one of the main results of this paper, and illustrates perturbative control over the structure of track function moments. They enable the calculation of up to the six point correlation functions in energy flow, matching the state of the art measured at the LHC in jet substructure. Our approach can be straightforwardly extended to compute higher moments of the track functions, as desired.

\section{Numerical Studies of Track Function Evolution}\label{sec:num}

In this section we numerically study the structure of the evolution equations for the track function moments. The goal of this section is two-fold. First, we show that $\Delta$ is sufficiently small in QCD, that corrections to DGLAP for the first three moments are effectively suppressed by (at least) an order in the perturbative expansion, allowing us to extend their RG evolution to NNLO. Second, we show that for the fourth moment and beyond, non-linearities in the evolution give rise to genuinly new behaviour beyond DGLAP.

\subsection{The Size of $\Delta$ in QCD and Extension to NNLO}\label{sec:delta}

We begin by studying the numerical impact of $\Delta$ for the first three-moments. The evolution of the first three central moments is constrained by shift symmetry to be of the form
\begin{align}\label{eq:structure_NNLO}
    \frac{\df}{\df\ln\mu^2}\Delta
    &=-\big[\gamma_{gg}(2)+\gamma_{qq}(2)\big]\Delta \, , \nn\\
    \frac{\df}{\df\ln\mu^2}\vec\sigma(2)
    &=-\hat\gamma(3)\vec\sigma(2)+\vec\gamma_{\Delta^2}\Delta^2 \, , \nn \\
    \frac{\df}{\df\ln\mu^2}\vec\sigma(3) 
    &=-\hat\gamma(4)\vec\sigma(3)+\hat\gamma_{\sigma_2\Delta}\vec\sigma(2)\Delta
    +\vec\gamma_{\Delta^3}\Delta^3 \,, 
\end{align}
where the evolution of $\Delta$ is fixed by DGLAP to all orders. For the second and third moment the evolution can be split into two parts: a linear term fixed by DGLAP and corrections proportional to powers of $\Delta$. Recall that $\Delta=T_q(1)-T_g(1)$, or more generally in the multi-flavor case is give by differences between the first moments of the track functions of different flavors. Since QCD final states at high energies are dominated by large numbers of nearly massless pions, the average values of the track functions are largely fixed by isospin, and hence satisfy $T_g(1)\simeq T_q(1)\simeq 2/3$,  and $\Delta \simeq 0$. Small corrections to this pictures give rise to $\Delta \ll 1$ in real world QCD. This suppression of $\Delta$, combined with the shift symmetry is particularly convenient, since it effectively suppresses the corrections to DGLAP by (at least) an order in the perturbative expansion. Indeed, we will see that this allows us to include the NNLO corrections to the DGLAP evolution while keeping the terms involving $\Delta$ at NLO. In our numerical studies we use the following initial conditions~\cite{Chang:2013iba},
\begin{alignat}{3}\label{eq:initial}
&T_g(1) = 0.624 \, , \qquad &&T_g(2) = 0.417 \, , \qquad &&T_g(3) = 0.293 \, , \nn \\
&T_q(1) = 0.611 \, , \qquad &&T_q(2) = 0.425 \, , \qquad &&T_q(3) = 0.319 \,,
\end{alignat}
at $\mu = 10$ GeV, and $\al_s(M_Z) = 0.116$ with $n_f=5$.

\begin{figure}
    \centering
    \includegraphics[scale=0.58]{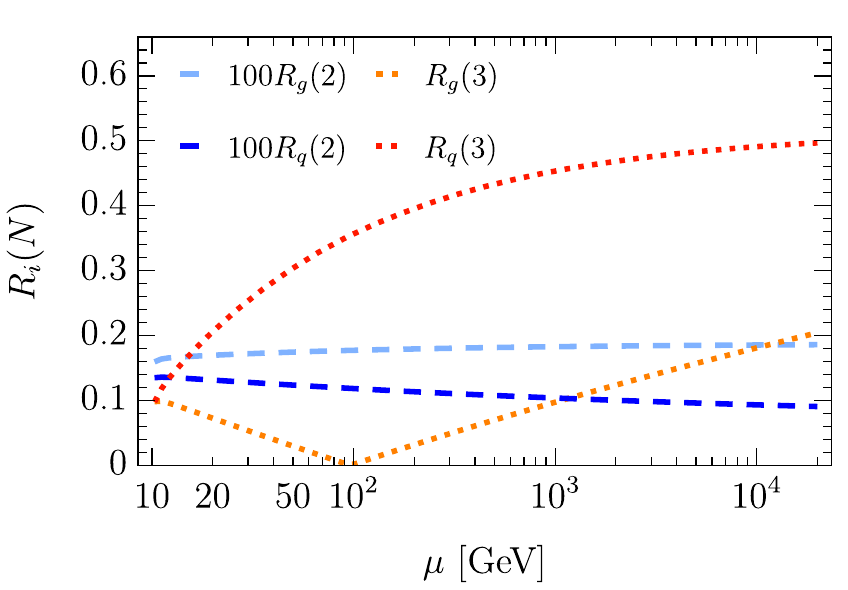}
    \caption{The ratio defined in eq.~\eqref{eq:deltannloratio} for the quark (darker) and gluon (lighter) second (blue dashed) and third (orange dotted) central moments as a function of the renormalization scale $\mu$. Note that the ratio for the second moment has been amplified by a factor 100 such that it is visible in this plot. The effect of $\Delta$ on the evolution of the second central moment is much smaller than for the higher moments because $\Delta$ appears only squared in the evolution for $\sigma(2)$, while for the other moments terms linear in $\Delta$ are also allowed.}
    \label{fig:deltavsnnlo}
\end{figure}

\begin{figure}
\begin{center}
\subfloat[]{
\includegraphics[scale=0.48]{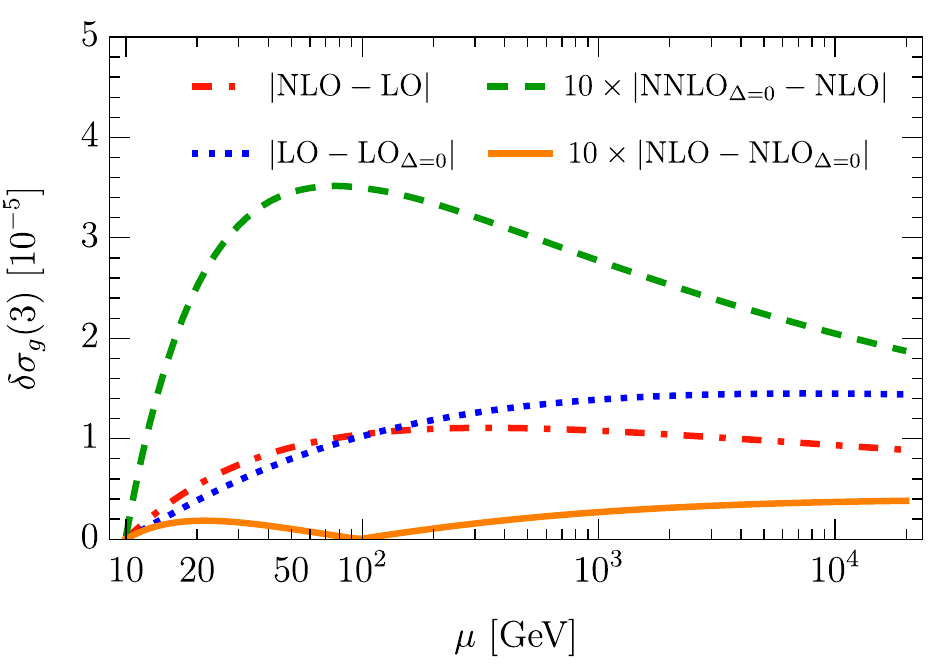}\label{fig:deltaatnlo_a}
}
\quad \subfloat[]{
\includegraphics[scale=0.48]{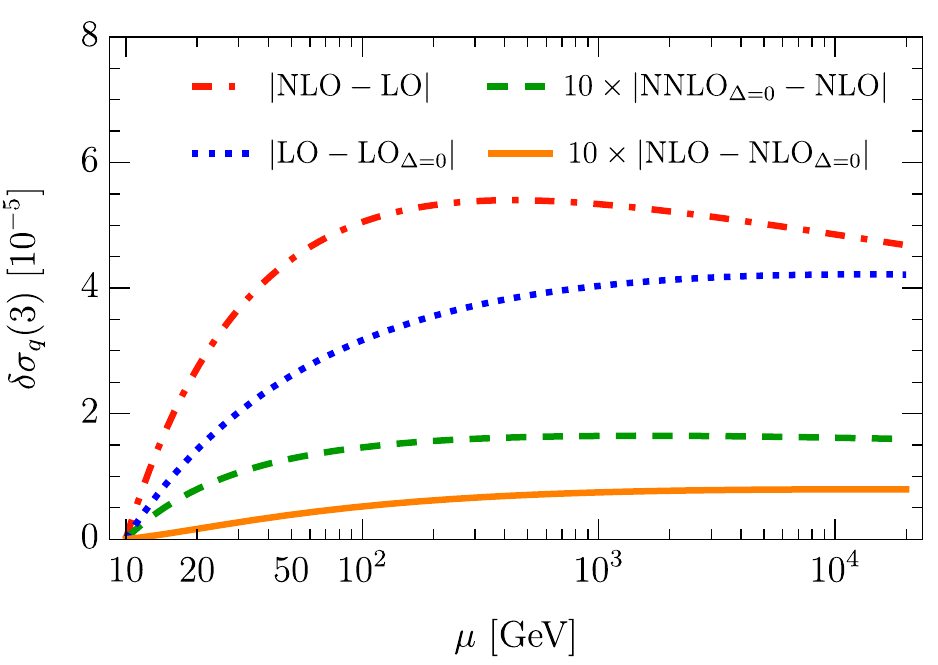}\label{fig:deltaatnlo_b}
}\qquad
\end{center}
\caption{The difference in renormalization group evolution for (a) $\sigma_g(3)$ and for (b) $\sigma_q(3)$ for the initial conditions in \eqref{eq:initial}. Shown are the effect of the $\Delta$ terms at LO (blue dotted), NLO (orange solid), the effect of the NLO evolution (red dot-dashed) and the NNLO evolution (green dashed). Note that two curves are multiplied by 10 for better visibility.}
\label{fig:deltaatnlo}
\end{figure}

To demonstrate that the effect of $\Delta$ on the evolution is much smaller than that of DGLAP, we study the following ratio
\begin{align}\label{eq:deltannloratio}
    R_i(n)&=\bigg|\frac
    {\sigma_i(n)|_{\text{NLO},\Delta=0}-\sigma_i(n)|_{\text{NLO}}}
    {\sigma_i(n)|_{\text{NNLO},\Delta=0}-\sigma_i(n)|_{\text{NLO}}}
    \bigg| \ .
\end{align}
In this ratio we compare the effect of including $\Delta$ with the effect of including the NNLO corrections to the DGLAP evolution. 
The notation $\si_i|_{\text{(N)NLO},\Delta=0}$ means setting the $\Delta$ terms in the (N)NLO evolution to zero, but \emph{not} in the  lower order terms of the evolution. We note that this ratio is scale dependent, and furthermore depends strongly on the value of $\Delta$. Since this ratio is meant to illustrate the approximate size, we have for simplicity kept the initial conditions the same for all scenarios, using the values in eq.~\eqref{eq:initial}. Figure \ref{fig:deltavsnnlo} shows this ratio for a range of values of $\mu$, which is much smaller than 1 for the second moment, as it only involves $\Delta^2$ terms. For the third moment, which involves terms linear in $\Delta$, the ratio is of order 1, indicating that the $\Delta$ terms at NLO are of the same size as the NNLO correction to the DGLAP evolution.  The (unknown) $\Delta$ terms at NNLO are of course much smaller. We further investigate the various contributions to the third moment in figure \ref{fig:deltaatnlo}. Here we show the size of the NLO evolution, the $\Delta$ term in the LO and NLO evolution, and the NNLO evolution (without $\Delta$ term) by taking appropriate differences, demonstrating that the $\Delta$ terms are effectively suppressed by one order in the perturbative expansion. The $\Delta$ terms at NNLO can therefore safely be neglected.

This allows us to immediately extend the evolution of the first three central moments of the track function to NNLO using known results for the timelike spin-$n$ anomalous dimensions \cite{Chen:2020uvt}. This simplification is quite convenient, as it allows us to immediately consider NNLO evolution for up to three-point correlators. For convenience, we provide the DGLAP anomalous dimensions for the first three moments up to NNLO in \App{app:splitting_moments}.

\subsection{Non-Linearities in the Fourth and Fifth Moments}\label{sec:nonlinear}

Although the evolution of the first three moments are DGLAP up to correction in $\Delta$, this is not the case for higher moments. This is because the evolution of higher moments can contain non-linear terms that are not proportional to $\Delta$ and are therefore not suppressed, even in a pure gluon theory. For example, the evolution of the fourth and fifth central moment is constrained by shift symmetry to be of the form
\begin{align}
    \frac{\df}{\df\ln\mu^2}\vec\sigma(4) 
    &=-\hat\gamma(5)\vec\sigma(4)
    +\hat\gamma_{\sigma_2\sigma_2}\big[\vec\sigma(2) \cdot \vec\sigma^T(2)\big] +\hat \gamma_{\sigma_3 \Delta} \vec \sigma(3) \Delta+ \hat \gamma_{\sigma_2 \Delta^2} \vec \sigma(2) \Delta^2+ \vec \gamma_{\Delta^4} \Delta^4\,,
     \\
    \frac{\df}{\df\ln\mu^2} \vec \sigma(5) &=-\hat \gamma(6) \vec \sigma(5)+ \hat \gamma_{\sigma_3 \sigma_2}\big[\vec \sigma(3)\cdot\vec \sigma^T(2)\big]\nn \\
    & \quad + \hat \gamma_{\sigma_4 \Delta} \vec \sigma(4) \Delta+ \hat \gamma_{\sigma_2^2 \Delta} \big[\vec \sigma(2)\cdot \vec\sigma^T(2)\big] \Delta+ \hat \gamma_{\sigma_3 \Delta^2} \vec \sigma(3) \Delta^2+ \hat \gamma_{\sigma_2 \Delta^3} \vec \sigma(2) \Delta^3+ \vec \gamma_{\Delta^5} \Delta^5\,.
\nn\end{align}
While the terms involving $\Delta$ are suppressed, the terms involving products of $\sigma(2)$ and $\sigma(3)$ are not. These non-linear terms are not constrained by DGLAP and require additional calculational techniques. Therefore extending the evolution of higher track function moments to NNLO is beyond the scope of this paper.

\begin{figure}
\begin{center}
\subfloat[]{
\includegraphics[scale=0.6]{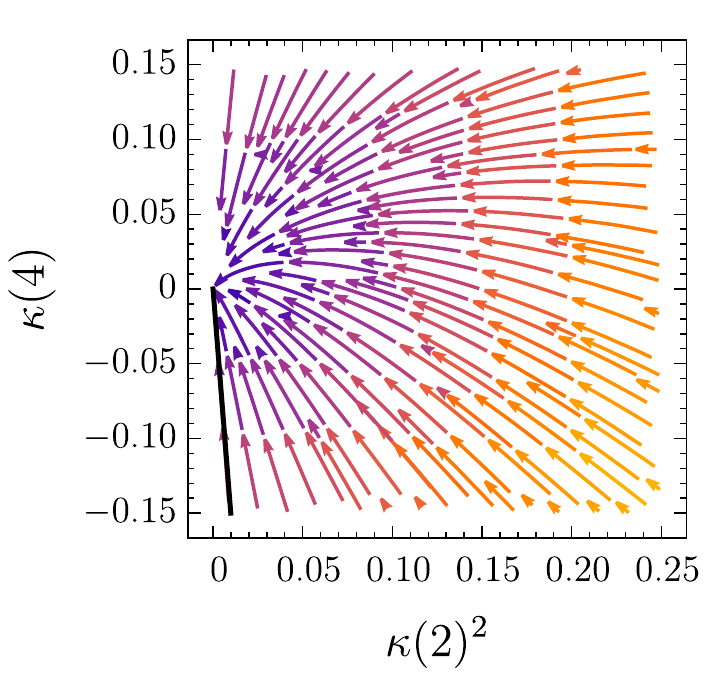}\label{fig:2dflow_a}
}
\qquad \subfloat[]{
\includegraphics[scale=0.6]{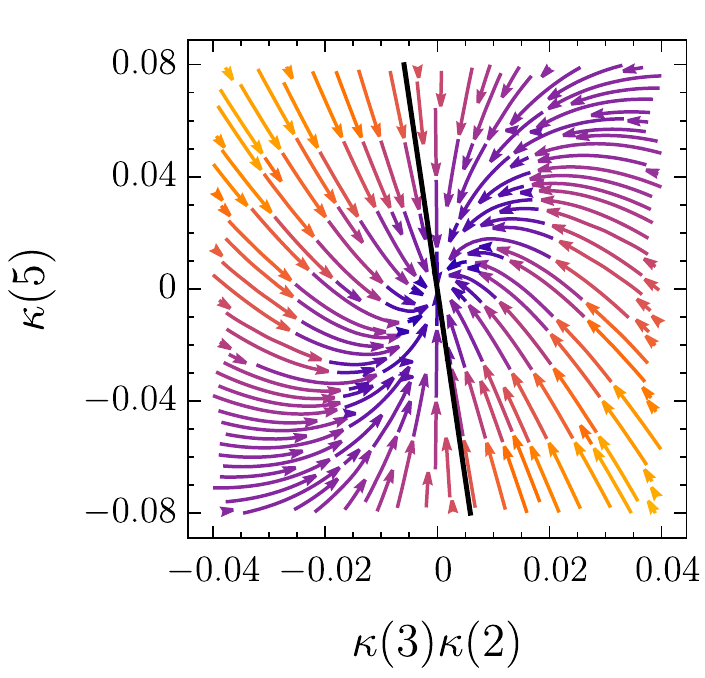}\label{fig:2dflow_b}
}\qquad
\end{center}
\caption{Renormalization group flow in pure Yang-Mills theory at fixed $\mu$ for (a) the fourth cumulant and (b) for the fifth cumulant. The arrows denote the direction of the derivatives with respect to $\ln\mu$ and their color reflects their strength. The black line indicates the eigenvector of the evolution equation.}
\label{fig:2dflow}
\end{figure}

Let us continue to study the effects of the non-linear terms in the evolution equations. For simplicity we consider the evolution of the fourth and fifth cumulant in pure Yang-Mills theory, where the evolution of these moments simplifies to
\begin{align}
    \frac{\df}{\df\ln\mu^2}\kappa(4)&=-\gamma_{gg}(5)\kappa(4)
    +\gamma_{\kappa_2\kappa_2} \kappa^2(2) \ , \nn\\
    \frac{\df}{\df\ln\mu^2}\kappa(5)&=-\gamma_{gg}(6)\kappa(5)
    +\gamma_{\kappa_3\kappa_2} \kappa(3)\kappa(2) \ .
\end{align}
These simplified expressions allow us to study the non-linearity of the evolution by means of a two-dimensional RG flow plot, shown in figure \ref{fig:2dflow}. This figure shows the RG flow for the fourth and fifth cumulant in the $\kappa(4)-\kappa^2(2)$ and $\kappa(5)-\kappa(3)\kappa(2)$ planes respectively. From these plots it is clear that there is a single fixed-point in the evolution at the origin, corresponding to the trivial fixed point where all cumulants vanish. In addition to this fixed point, the flow lines are attracted to a common valley before flowing to the fixed point. Note that the range of the axes on these plots are somewhat arbitrary, as the figure is invariant under a simultaneous rescaling of both axes.

While it is clear that the trivial fixed point is an attractive fixed point, these plots give interesting insight into the behavior of the track function. For example, we can consider a Gaussian track function for which all higher cumulants vanish. In this case, the track function will first generate a non-zero value of $\kappa(4)$ through the non-linear mixing, after which the DGLAP anomalous dimensions drive it back to zero. In this case, which is a good approximation to real world QCD, the mixing anomalous dimensions dominate the behavior of the track function evolution. Since physically the distribution must eventually collapse to a delta function under RG evolution, this suggests that there should be a positivity bound on $\gamma_{\kappa_2 \kappa_2}$. This provides further evidence that it may have a direct interpretation as an anomalous dimension of some generalized lightray operator, and it would be interesting to understand this better.

The RG flow of the fifth cumulant, $\kappa(5)$, is interesting in that it illustrates the structure of odd moments. The RG of the track functions preserves symmetry/anti-symmetry properties under RG flow. This is manifest in the $\kappa(5) \to -\kappa(5)$, $\kappa(3)\to -\kappa(3)$ symmetry of the RG flow in the figure.   For higher moments, additional non-linear terms in the evolution appear and a visualization of the RG flow can only be realized in higher-dimensional RG flow plots.

Due to the dominance of mixing terms beyond the third moment, we are not immediately able to extend our calculation to NNLO. While the complete calculation of the NNLO evolution of higher moments is beyond the scope of this paper, we briefly comment on what would be required to do so. The constraints from shift symmetry hold to all orders in perturbation theory. Focusing on pure Yang-Mills theory for simplicity, one can show that to all orders in perturbation theory the fourth moment takes the form
\begin{align}
\mu \frac{d}{d\mu}T(4)&=-\gamma(5) T(4)-4(\gamma(4)-\gamma(5))T(1)T(3)+ (\gamma_{1\to 4}-6\gamma(3)+8\gamma(4)-3\gamma(5)) T(2)T(2) \nn \\
&\quad -2(\gamma_{1\to 4} -3\gamma(3)+2\gamma(4))T(1)T(1)T(2)
+\gamma_{1\to 4} T(1)T(1)T(1)T(1)
\end{align}
Here we see that only one anomalous dimension, $\gamma_{1\to 4}$, beyond the standard DGLAP anomalous dimension, appears. Interestingly, this particular contribution does not involve any soft singularities, since it has one energy weighting on each parton. Its calculation is therefore much simpler than calculations of the NNLO DGLAP kernels. It could be computed, for example, using the known $1\to4$ splitting functions \cite{DelDuca:2019ggv,DelDuca:2020vst}.

\section{Conclusions}\label{sec:conc}

Track functions characterize the fluctuations in the fragmentation process of quarks and gluons into charged hadrons (or some other subset of hadrons), and its moments are essential for the description of track-based measurements of higher-point correlation functions in jet substructure. Although they are fundamentally non-perturbative objects, the track function evolution is perturbative and exhibits interesting renormalization group structure involving mixings between different moments.

In this paper we have derived the all-orders structure of the RG for the moments of track functions, using the action of energy conservation as a shift symmetry. This highlights the remarkably constrained structure of the evolution, implying that the RG can be expressed in terms of cumulants (or equivalently, central moments), and differences of first moments. 

We performed an explicit calculation of the first six moments of the quark and gluon track functions in QCD. At the fourth moment and beyond one finds interesting RG flows describing the mixing with products of cumulants, for example between $\kappa(4)$ and $(\kappa(2))^2$. We studied the structure of these RG flows, finding that these mixing terms dominate the evolution. These higher cumulants of the track functions therefore probe evolution in the fragmentation process that goes beyond the standard DGLAP evolution, and it would be interesting to better understand the structure of these mixing terms in terms of anomalous dimensions of the underlying field theory, and study them experimentally.

Finally, we showed that for the first three moments cumulants of the track function, shift symmetry constrains any evolution beyond DGLAP to be proportional to $\Delta$. For track-based measurements in QCD, $\Delta \ll 1$, making the corrections proportional to $\Delta$ suppressed by an effective order in the perturbative expansion. This allows us to extend the evolution to NNLO, enabling up to three-point correlators to be studied on tracks at this order. We also outlined the missing ingredients for a similar extension to NNLO beyond the third moment, where genuinely new ingredients are required.

Although we have primarily focused in this paper on the experimental utility of track functions, we believe that better understanding the evolution of the moments of the track functions could be of more formal theoretical interest. The DGLAP anomalous dimensions have a deep connection to the twist-2 operators of the theory, which has recently recieved renewed attention in the study of lightray operators in CFTs \cite{Kravchuk:2018htv}. Track functions are another class of intrinsically Lorentzian observables, that probe features of the theory beyond the leading twist trajectory. It would be interesting if they could be put on a similarly firm theoretical footing, and if one could more precisely understand what features of the theory they are probing, and how they are related to its operator content.


Our results allow the calculation of up-to six point energy correlators on tracks, which have recently been investigated with CMS open data~\cite{Komiske:2022enw,Chen:2022swd,Lee:2022ige} providing a view on the hadronization transition, non-Gaussianities and quantum scaling dimensions. The three-point energy correlator has also been proposed as a new way to extract the top quark mass~\cite{Holguin:2022epo}, with the potential to reduce the theoretical uncertainty, particularly from nonperturbative effects. The angular resolution offered by tracks is essential to carry out these measurements. This is also the case for the azimuthal decorrelation in vector-boson plus jet production~\cite{Chien:2020hzh}, which however requires knowledge of (the evolution of) the full track function. In conclusion, we believe that our work will be of significant interest for precision studies at the LHC, and we look forward to their application in phenomenology in the near future.

\acknowledgments

We thank Solange Schrijnder van Velzen for collaboration in the early stages of this work. We thank Hao Chen, Patrick Komiske, Jesse Thaler, Tong Zhi Yang and Xiao Yuan Zhang for helpful discussions. 
M.J.~is supported by the NWO projectruimte 680-91-122. 
Y.L. and H.X.Z.~are supported by the National Natural Science
Foundation of China under contract No.~11975200.
I.M.~is supported by start up funds from Yale University.
W.W.~is supported by the D-ITP consortium, a program of NWO that is funded by the Dutch Ministry of Education, Culture and Science (OCW).


\appendix

\section{Splitting Function Calculation for Pure Yang-Mills}\label{app:yang_mills}

We will now discuss how the anomalous dimensions for pure Yang-Mills can be calculated using the approach of sec.~\ref{sec:calc_split}. We will employ the notation for the anomalous dimensions in sec.~\ref{sec:glu}.

In our method we will consider $\ga(a,b,c)$ with $a, b, c>0$, which allows us to avoid soft singularities in the integrations. For pure Yang-Mills this is sufficient, since $\ga(a,0,0) = \tfrac13 \ga_a$ is the known anomalous dimension of the fragmentation function, and we can fix $\ga(a,b,0)$ using 
\begin{align}
   \sum_{c=0}^b \ga(a,b-c,c) &=
   \int_0^1\! \df z_1\,\df z_2\, \df z_3\, \de(1 - z_1 - z_2 - z_3) P(\{z_i\})   
     \sum_{c=0}^b
      \begin{pmatrix} & a+b & \\ a & b-c & c \end{pmatrix} z_1^a z_2^{b-c} z_3^c 
  \nn \\ &=
    \frac13 \sum_{c=0}^b \begin{pmatrix} & a+b & \\ a & b-c & c\end{pmatrix} (-1)^c \ga_{a+c}
\,.\end{align}
This follows, because \emph{under the integral} we can make the replacement
\begin{align}
  \sum_{c=0}^b  \begin{pmatrix} & a\!+\!b & \\ a & b\!-\!c & c \end{pmatrix} z_1^a z_2^{b-c} z_3^c 
  =  \begin{pmatrix} a\!+\!b \\ b \end{pmatrix} z_1^a (z_2\!+\!z_3)^b   
  =  \begin{pmatrix} a\!+\!b \\ b \end{pmatrix} z_1^a (1\!-\!z_1)^b
  =  \sum_{c=0}^b  \begin{pmatrix} & a\!+\!b & \\ a & b\!-\!c & c \end{pmatrix} (-1)^c z_1^{a+c}
\,.\end{align}

As discussed in sec.~\ref{sec:calc_split}, we will extract the track function evolution from the jet function $J(s,x)$ differential in the total invariant mass $s$ of the jet and the track fraction $x$, by integrating the collinear splitting amplitudes. Because the measurement of $x$ only receives contributes from collinear radiation (contributions from soft radiation are power suppressed), consistency of factorization in SCET implies that this jet function must have the same anomalous dimension as the well-known jet function that is only differential in the invariant mass $s$~\cite{Becher:2010pd}. After this renormalization, the remaining divergences must be IR in nature and absorbed by the track functions, as encoded by the matching relation
\begin{align}\label{T_matching}
 J(s,x;\mu) = \sum_N  \biggl[\prod_{i=1}^N \int_0^1\! \df z_i \biggr] \de\Bigl(1 - \sum_{i=1}^N z_i\Bigr) \mathcal{J}_{1 \to N}(s,\{z_i\},\mu) \int \biggl[\prod_{i=1}^N\df x_i\, T(x_i,\mu)\biggr] \de\Bigl(x - \sum_{i=1}^N z_i x_i\Bigr)\,.
\end{align}
Note that this matching is between \emph{renormalized} quantities, as is standard. Since we are working in a pure gluon theory, we have removed all flavor labels. Note that the matching coefficients $\mathcal{J}$ are IR finite. The IR poles in the track functions follow from the inverse of \eqref{eq:poles}, which reads
\begin{align}
    \vec{\mathbf{T}}_n(\mu)=\left\{
        1-a_s(\mu)\frac{\widehat{R}_n^{(1)}}{\epsilon}+\frac{1}{2}a_s^2(\mu)\left(-\frac{\widehat{R}_n^{(2)}}{\epsilon}+\frac{\widehat{R}_n^{(1)}\widehat{R}_n^{(1)}+\beta_0\widehat{R}_n^{(1)}}{\epsilon^2}\right)+O(a_s^3)
    \right\}
    \vec{\mathbf{T}}_n^{(0)}
\,.\end{align}
At order $\alpha_s^2$, we get the contribution $\mathcal{J}^{(0)} \otimes T^{(2)} = \de(s)\,T^{(2)}(x)$ in \eqref{T_matching}, which gives us the desired IR poles of the renormalized track function, from which we can infer the UV poles $\widehat{R}_n^{(2)}$ and anomalous dimension. This also tells us that we can restrict our attention to the coefficient of the $\delta(s)$ term in \eq{T_matching}. The cross term involving $\mathcal{J}^{(1)}$ and $T^{(1)}$ can be taken into account, using the matching coefficients for fragmenting jet functions~\cite{Jain:2011xz}, which are the same at this order (since the momentum fraction of the other branch is simply $1-z$). Finally, the  $\mathcal{J}^{(2)}$ contribution can be ignored, since it does not contain any poles.

We will now describe the calculation of the jet function $J(s,z)$ in some detail. Since we restrict to $a,b,c>0$, only the double real contribution needs to be included,
\begin{align}
  J(s,x) &= \frac{1}{6} \int\! \df \Phi_3^c\, \de(s - s_{123})\, \sigma_3^{c} \int \biggl[\, \prod_{i=1}^3 \df x_i\, T^{(0)}(x_i)\biggr] \de\Bigl(x - \sum_{i=1}^3 z_i x_i\Bigr) + \dots
\,.\end{align}
Here  $\df \Phi_3^c$ the three-body collinear phase space~\cite{Gehrmann-DeRidder:1997fom} for non-identical particles (hence the $\tfrac16$) 
\begin{align}
\df \Phi_3^c &= \df s_{123}\, \df s_{12}\,\df s_{13}\, \df s_{23}\,  \de(s_{123}-s_{12}-s_{13}-s_{23})
 \df z_1\, \df z_2\, \df z_3\,   \de(1-z_1-z_2-z_3)
 \nn \\ & \quad \times 
 \frac{4\Theta(-\De)(-\De)^{-\frac12-\eps}}{(4\pi)^{5-2\eps} \Ga(1-2\eps)}
\,,\end{align}
with $z_i$ the momentum fraction of parton in $i$, $s_{ij}$ the invariant mass of partons $i$ and $j$ and 
\begin{align}
   \De &=  (z_3 s_{12} - z_1 s_{23}- z_2 s_{13})^2 - 4z_1z_2s_{13}s_{23}
\,.\end{align}
The squared collinear matrix element $\sigma_3^{c}$ describing the $g \to ggg$ splitting is~\cite{Campbell:1997hg,Catani:1998nv}
\begin{align}
  \si_3^c &= \Big(\frac{\mu^2 e^{\ga_E}}{4\pi}\Big)^{2\eps}\, \frac{4g^4 C_A^2}{s_{123}^2} 
  \bigg\{\frac{(1-\eps)}{4s_{12}^2} \Bigl(2 \frac{z_1s_{23} - z_2 s_{13}}{z_1 + z_2} + \frac{z_1 - z_2}{z_1 + z_2} s_{12}\Bigr)^2 + \frac34(1-\eps) + \frac{s_{123}}{s_{12}}\biggl[4 \frac{z_1 z_2-1}{1-z_3} 
  \nn \\ & \quad  
  + \frac{z_1 z_2 - 2}{z_3} + \frac32 + \frac52 z_3 
  + \frac{(1- z_3(1-z_3))^2}{z_3 z_1(1-z_1)}\biggr] + \frac{s_{123}^2}{s_{12}s_{13}} \biggr[ \frac{z_1 z_2(1-z_2)(1-2z_3)}{z_3(1-z_3)} + z_2 z_3 - 2 
  \nn \\ & \quad  
  + \frac{z_1(1+2z_1)}{2}   + \frac{1+2z_1(1+z_1)}{2(1-z_2)(1-z_3)} + \frac{1-2z_1(1-z_1)}{2z_2z_3}\biggr]\biggr\} + \text{5 permutations}.
\end{align}
The integral over $s_{ij}$ can be carried out analytically using the results in the appendix of Ref.~\cite{Kosower:2003np}.
Since we only need the coefficient of $\delta(s)$, we can restrict ourselves to the first term in the plus expansion $s^{-1-2\eps} = -1/(2\eps)\, \delta(s) + \dots$. By taking the $n$-th moment, we can rewrite
\begin{align}
  &\int\! \df x\, x^n \prod_{i=1}^3\biggl[\int\!\df x_i\, T^{(0)}(x_i)\biggr]  \de\Bigl(x - z_1 x_1 - z_2 x_2 - z_3 x_3) 
  \nn \\ & \quad
  = \sum_{a+b+c=n} \begin{pmatrix} & n & \\ a & b & c \end{pmatrix} z_1^a z_2^b z_3^c\, T^{(0)}(a) T^{(0)}(b) T^{(0)}(c)
\,.\end{align}
Because we restricted our attention to those terms with $a,b,c>0$ there are no soft singularities, allowing us to first expand in $\eps$ and then integrate over $z_i$. 

Finally, to subtract the contribution involving $\mathcal{J}^{(1)}$ and $T^{(1)}$, we need to transform \eq{T_matching} to moment space. Keeping only the $\de(s)$ term, 
\begin{align} 
  J(s,n) &= a_s^2 \de(s) \biggl[ T^{(2)}(n) + 2 \sum_{a+b=n} \begin{pmatrix} n \\ a \end{pmatrix} 
  \mathcal{J}^{(1)}_\delta(a,b) T^{(0)}(a) T^{(1)}(b) \biggr] + \dots
\,,\end{align}
where $a_s = \alpha_s/(4\pi)$ and (assuming $a,b,c>0$)
\begin{align}
  T^{(1)}(b) &= - \frac{1}{\eps} \sum_{c=0}^b  T^{(0)}(b-c) T^{(0)}(c) \int_0^1\!\df z\, z^b (1-z)^c p_{gg}(z)\,, \nn \\
  \mathcal{J}^{(1)}(a,b) &=  \int_0^1\!\df z\, z^a (1-z)^b\, \ln[z(1-z)]\,p_{gg}(z)\,, \nn \\
  p_{gg}(z) &= 2 C_A \Bigl[\frac{z}{1-z} + \frac{1-z}{z} + z(1-z)\Bigr]
\,.\end{align}
Using this approach we have determined the unknown anomalous dimensions in pure Yang-Mills up to the ninth moment (see \eqref{eq:ga_rel})
\begin{align}
  \ga_{42}^{(2)} &= C_A^2\Bigl(\frac{47613060961}{22226400} - \frac{2321 \pi^2}{9} + 360 \zeta_3\Bigr)
  \,,\nn\\
  \ga_{62}^{(2)} &=  C_A^2 \Bigl(\frac{6322515311879}{1440747000} - \frac{777388 \pi^2}{1485} + 672\zeta_3\Bigr)
  \,,\nn\\
  \ga_{72}^{(2)} &= C_A^2 \Bigl(\frac{22916518522033}{18489586500} - \frac{182096 \pi^2}{1155} + 288\zeta_3\Bigr)
\,,\end{align}
and it is easy to obtain results for higher moments.

\section{Moments of Timelike Splitting Functions}\label{app:splitting_moments}

The timelike splitting functions are
\begin{align}
P_{ij}(z)=\sum_{L=0}^\infty a_s^{L+1}P_{ij}^{(L)}(z)\ ,
\end{align}
where $a_s = \alpha_s/(4\pi)$. The Mellin moments of timelike splitting functions are
\begin{align}
\gamma_{ij}^{(L)}(k)=-\int_0^1\df z\ z^{k-1}P_{ij}^{(L)}(z)\ .
\end{align}
Note that this is shifted by one from the definition of the moments of the track function in \eq{eq:T_mom}.
All the results of $P_{ij}(z)$ up to order-$a_s^3$ are e.g. listed in the ancillary file, ``PT.txt'', of~\cite{Chen:2020uvt}, and $P_{ij}(z)$ corresponds to PT[``$ij$''] in that file. At LO, $P_{\bar{q}q}^{(0)}$ and $P_{Qq}^{(0)}$ vanish while the non-vanishing moments up to the 7th moment are given by
\begin{alignat}{4}
&\gamma_{gg}^{(0)}(2)=\frac{4}{3}n_f T_F \ , \quad
&&\gamma_{qg}^{(0)}(2)=-\frac{2}{3}T_F \ , \quad
&&\gamma_{gq}^{(0)}(2)=-\frac{8}{3}C_F \ , \quad
&&\gamma_{qq}^{(0)}(2)=\frac{8}{3}C_F \ ,
\nonumber\\
&\gamma_{gg}^{(0)}(3)=\frac{4}{3}n_f T_F +\frac{14}{5}C_A\ , \quad
&&\gamma_{qg}^{(0)}(3)=-\frac{7}{15}T_F \ , \quad
&&\gamma_{gq}^{(0)}(3)=-\frac{7}{6}C_F \ , \quad
&&\gamma_{qq}^{(0)}(3)=\frac{25}{6}C_F \ ,
\nonumber\\
&\gamma_{gg}^{(0)}(4)=\frac{4}{3}n_f T_F +\frac{21}{5}C_A\ , \quad
&&\gamma_{qg}^{(0)}(4)=-\frac{11}{30}T_F \ , \quad
&&\gamma_{gq}^{(0)}(4)=-\frac{11}{15}C_F \ , \quad
&&\gamma_{qq}^{(0)}(4)=\frac{157}{30}C_F \ ,
\nonumber\\
&\gamma_{gg}^{(0)}(5)=\frac{4}{3}n_f T_F +\frac{181}{35}C_A\ , \quad
&&\gamma_{qg}^{(0)}(5)=-\frac{32}{105}T_F \ , \quad
&&\gamma_{gq}^{(0)}(5)=-\frac{8}{15}C_F \ , \quad
&&\gamma_{qq}^{(0)}(5)=\frac{91}{15}C_F \ ,
\nonumber\\
&\gamma_{gg}^{(0)}(6)=\frac{4}{3}n_f T_F +\frac{83}{14}C_A\ , \quad
&&\gamma_{qg}^{(0)}(6)=-\frac{11}{42}T_F \ , \quad
&&\gamma_{gq}^{(0)}(6)=-\frac{44}{105}C_F \ , \quad
&&\gamma_{qq}^{(0)}(6)=\frac{709}{105}C_F \ ,
\nonumber\\
&\gamma_{gg}^{(0)}(7)=\frac{4}{3}n_f T_F +\frac{4129}{630}C_A\ , \quad
&&\gamma_{qg}^{(0)}(7)=-\frac{29}{126}T_F \ , \quad
&&\gamma_{gq}^{(0)}(7)=-\frac{29}{84}C_F \ , \quad
&&\gamma_{qq}^{(0)}(7)=\frac{1027}{140}C_F \ .
\end{alignat}
At NLO, 
\begin{align}
\gamma_{gg}^{(1)}(2)&=
n_f T_F \bigg[\left(\frac{200}{27}-\frac{16 \pi^2}{9}\right)C_A 
+\frac{260}{27}C_F\bigg] \ ,
\nonumber\\
\gamma_{gg}^{(1)}(3)&=
n_fT_F \bigg[\left(\frac{3803}{675}-\frac{16 \pi ^2}{9}\right)C_A +\frac{12839}{2700}C_F\bigg] 
+\left(\frac{2158}{675}+\frac{26 \pi ^2}{45}-8\zeta_3\right)C_A^2 \ ,
\nonumber\\
\gamma_{gg}^{(1)}(4)&=
n_f T_F\bigg[\left(\frac{2273}{675}-\frac{16 \pi ^2}{9}\right)C_A
+\frac{57287}{13500}C_F\bigg] 
+\left(\frac{90047}{1500}-\frac{28 \pi^2}{5}\right)C_A^2 \ ,
\nonumber\\
\gamma_{gg}^{(1)}(5)&=
n_f T_F\bigg[\left(-\frac{16 \pi ^2}{9}+\frac{52798}{33075}\right)C_A
+\frac{680132}{165375}C_F\bigg] 
+\left(\frac{4706626}{165375}-\frac{316 \pi^2}{315}-8\zeta_3\right)C_A^2 \ , 
\nonumber\\
\gamma_{gg}^{(1)}(6)&=
n_f T_F\bigg[\left(-\frac{16\pi ^2}{9}+\frac{2071}{13230}\right)C_A 
+\frac{940633}{231525}C_F\bigg] 
+\left(\frac{13375435}{148176}-\frac{166 \pi ^2}{21}\right)C_A^2 \, 
\nonumber\\
\gamma_{gg}^{(1)}(7)&=
n_f T_F\bigg[\left(\!-\!\frac{1262143}{1190700}\!-\!\frac{16\pi^2}{9}\right)C_A
\!+\!\frac{10772855}{2667168}C_F\bigg]
+\left(\frac{2907487777}{66679200}\!-\!\frac{1819\pi^2}{945}\!-\!8\zeta_3\right)C_A^2 , 
\nonumber\\
&
\nonumber\\
\gamma_{qg}^{(1)}(2)&=
T_F\bigg[\left(\frac{8 \pi ^2}{9}-\frac{100}{27}\right)C_A
-\frac{130}{27}C_F\bigg] \ , 
\nonumber\\
\gamma_{qg}^{(1)}(3)&=
T_F \bigg[\left(\frac{619}{2700}+\frac{14 \pi ^2}{45}\right)C_A
-\frac{833}{216}C_F-\frac{8}{25}n_f T_F\bigg] \ ,
\nonumber\\
\gamma_{qg}^{(1)}(4)&=
T_F\bigg[\left(\frac{22 \pi ^2}{45}-\frac{60391}{27000}\right)C_A
-\frac{166729}{54000}C_F -\frac{12}{25}n_f T_F\bigg] \ ,
\nonumber\\
\gamma_{qg}^{(1)}(5)&=
T_F\bigg[\left(\frac{1999}{18375}+\frac{64 \pi ^2}{315}\right) C_A 
-\frac{19792}{7875}C_F-\frac{2048}{3675}n_f T_F\bigg]\ , 
\nonumber\\
\gamma_{qg}^{(1)}(6)&=
T_F\bigg[\left(\frac{22 \pi^2}{63}-\frac{1249361}{740880}\right) C_A
 -  \frac{427303}{205800}C_F  - \frac{436}{735}n_f T_F\bigg] \ , 
\nonumber\\
\gamma_{qg}^{(1)}(7)&=
T_F\bigg[\left(\frac{29 \pi ^2}{189}-\frac{674773}{22226400}\right)C_A 
-\frac{77139049}{44452800}C_F -\frac{36158}{59535} n_f T_F\bigg] \ , 
\nonumber\\
&
\nonumber\\
\gamma_{gq}^{(1)}(2)&=
\left(\frac{32\pi^2}{9}-\frac{568}{27}\right)C_F^2 
-\frac{376}{27}C_A C_F \ ,
\nonumber\\
\gamma_{gq}^{(1)}(3)&=
\left(\frac{14\pi^2}{9}-\frac{2977}{432}\right)C_F^2 
+\left(-\frac{39451}{5400}-\frac{7\pi^2}{9}\right)C_A C_F \ ,
\nonumber\\
\gamma_{gq}^{(1)}(4)&=
\left(\frac{44\pi^2}{45}-\frac{104389}{27000}\right)C_F^2
-\frac{142591}{13500}C_A C_F \ ,
\nonumber\\
\gamma_{gq}^{(1)}(5)&=
\left(\frac{32\pi^2}{45}-\frac{9374}{3375}\right)C_F^2
+\left(-\frac{2882863}{661500}-\frac{16\pi^2}{45}\right)C_A C_F \ ,
\nonumber\\
\gamma_{gq}^{(1)}(6)&=
\left(\frac{176\pi^2}{315}-\frac{2626061}{1157625}\right)C_F^2
-\frac{948127}{154350}C_A C_F \ ,
\nonumber\\
\gamma_{gq}^{(1)}(7)&=
\left(\frac{29\pi^2}{63}-\frac{19635271}{9878400}\right)C_F^2
+\left(-\frac{358501999}{133358400}-\frac{29\pi^2}{126}\right)C_A C_F \ ,
\nonumber\\
&
\nonumber\\
\gamma_{qq}^{(1)}(2)&=
\left(\!-\!\frac{175}{27}\!+\!\frac{2\pi^2}{9}\!-\!8\zeta_3\right)C_F^2 
+\left(\frac{1495}{54}\!-\!\frac{17\pi^2}{9}\!+\!4\zeta_3\right)C_A C_F
+\left(\frac{64}{27}\!-\!\frac{128}{27}n_f\right)C_F T_F \ ,
\nonumber\\
\gamma_{qq}^{(1)}(3)&=
\left(\frac{989}{432}\!-\!\frac{7\pi^2}{9}\!-\!8\zeta_3\right)C_F^2
+\left(\frac{16673}{432}\!-\!\frac{43\pi^2}{18}\!+\!4\zeta_3\right)C_A C_F
+\left(\frac{4391}{5400}\!-\!\frac{415}{54}n_f\right)C_F T_F \ ,
\nonumber\\
\gamma_{qq}^{(1)}(4)&=
\left(\frac{55553}{6000}-\frac{67\pi^2}{45}-8\zeta_3\right)C_F^2
+\left(\frac{2495453}{54000}-\frac{247\pi^2}{90}+4\zeta_3\right)C_A C_F
\nonumber\\
&\qquad+\left(\frac{11867}{27000}-\frac{13271}{1350}n_f\right)C_F T_F \ ,
\nonumber\\
\gamma_{qq}^{(1)}(5)&=
\left(\frac{100669}{6750}-\frac{92\pi^2}{45}-8\zeta_3\right)C_F^2 
+\left(\frac{156421}{3000}-\frac{136\pi^2}{45}+4\zeta_3\right)C_A C_F 
\nonumber\\
&\qquad+\bigg(\frac{46516}{165375}-\frac{7783}{675}n_f\bigg)C_F T_F \ ,
\nonumber\\
\gamma_{qq}^{(1)}(6)&=
\left(\frac{363875}{18522}-\frac{788\pi^2}{315}-8\zeta_3\right)C_F^2
+\left(\frac{176024953}{3087000}-\frac{1024\pi^2}{315}+4\zeta_3\right)C_A C_F 
\nonumber\\
&\qquad+\left(\frac{3649}{18522}-\frac{428119}{33075}n_f\right)C_F T_F \ ,
\nonumber\\
\gamma_{qq}^{(1)}(7)&=
\left(\frac{234152309}{9878400}-\frac{607\pi^2}{210}-8\zeta_3\right)C_F^2
+\left(\frac{9065721869}{148176000}-\frac{1447\pi^2}{420}+4\zeta_3\right)C_A C_F
\nonumber\\
&\qquad+\left(\frac{779767}{5334336}-\frac{3745727}{264600}n_f\right)C_F T_F\ , 
\nonumber\\
&
\nonumber\\
\gamma_{\bar{q}q}^{(1)}(2)&=
\left(-\frac{743}{54}+\frac{17\pi^2}{9}-4\zeta_3\right)\frac{C_F}{N_c}
+\frac{64}{27} C_F T_F \ ,
\nonumber\\
\gamma_{\bar{q}q}^{(1)}(3)&=
\left(\frac{8113}{432}-\frac{43\pi^2}{18}+4\zeta_3\right)\frac{C_F}{N_c}
+\frac{4391}{5400} C_F T_F \ ,
\nonumber\\
\gamma_{\bar{q}q}^{(1)}(4)&=
\left(-\frac{1202893}{54000}+\frac{247\pi^2}{90}-4\zeta_3\right)\frac{C_F}{N_c}
+\frac{11867}{27000} C_F T_F \ ,
\nonumber\\
\gamma_{\bar{q}q}^{(1)}(5)&=
\left(\frac{675559}{27000}-\frac{136\pi^2}{45}+4\zeta_3\right)\frac{C_F}{N_c}
+\frac{46516}{165375} C_F T_F \ ,
\nonumber\\
\gamma_{\bar{q}q}^{(1)}(6)&=
\left(-\frac{252598609}{9261000}+\frac{1024\pi^2}{315}-4\zeta_3\right)\frac{C_F}{N_c}
+\frac{3649}{18522} C_F T_F \ ,
\nonumber\\
\gamma_{\bar{q}q}^{(1)}(7)&= 
\left(\frac{1442001293}{49392000}-\frac{1447\pi^2}{420}  +  4\zeta_3\right)\frac{C_F}{N_c} 
+\frac{779767}{5334336} C_F T_F\ ,
\end{align}
For $Q\neq q$ we have $\gamma_{Qq}=\gamma_{\bar{Q}q}$ and up to the 7th moment we have
\begin{alignat}{3}
 &\gamma_{Qq}^{(1)}(2)=\frac{64}{27}C_FT_F \ , \qquad
&&\gamma_{Qq}^{(1)}(3)=\frac{4391}{5400}C_FT_F \ , \qquad
&&\gamma_{Qq}^{(1)}(4)=\frac{11867}{27000}C_F T_F \ ,
\nonumber\\
 &\gamma_{Qq}^{(1)}(5)=\frac{46516}{165375}C_F T_F \ , \qquad
&&\gamma_{Qq}^{(1)}(6)=\frac{3649}{18522}C_F T_F \ , \qquad
&&\gamma_{Qq}^{(1)}(7)=\frac{779767}{5334336} C_F T_F \ .
\end{alignat}

For the EEC evolution to NNLL, we need the $N=3$ moment at LO, NLO and NNLO, which can be obtained from refs.~\cite{Mitov:2006wy,Mitov:2006ic,Moch:2007tx,Almasy:2011eq}. (Note that we include the pure singlet term in the $qq$ element.) At NNLO, we have
\begin{align}
\gamma_{gg}^{(2)}(2)&=
n_f T_F \bigg[\left(-\frac{256\zeta_3}{9}
+\frac{12464}{243}-\frac{2132 \pi^2}{81}+\frac{80 \pi^4}{27}\right)C_A^2 
\nonumber\\
&\quad\qquad
+\left(\frac{112\zeta_3}{9}+\frac{5362}{243}-\frac{760 \pi^2}{81}\right)C_A C_F +\left(-\frac{64\zeta_3}{9}+\frac{21140}{243}-\frac{352 \pi^2}{81}\right)C_F^2\bigg]
\nonumber\\
&\quad+n_f^2 T_F^2 \bigg[\left(-\frac{256\zeta_3}{9}-\frac{8}{27}
+\frac{320 \pi^2}{81}\right)C_A +\left(\frac{164}{9}
-\frac{256 \pi^2}{81}\right) C_F\bigg] \ , 
\nonumber\\
\gamma_{gg}^{(2)}(3)&=
\left(-\frac{23702\zeta_3}{225}+\frac{32 \pi^2\zeta_3}{3}+96\zeta_5
-\frac{5819653}{486000}+\frac{33179 \pi^2}{3375}-\frac{1283 \pi^4}{675}\right)C_A^3 
\nonumber\\
&\quad+n_f T_F \bigg[\left(\frac{478\zeta_3}{9}-\frac{12230737}{972000}
-\frac{51269 \pi^2}{1620}+\frac{104 \pi^4}{45}\right)C_A^2 
\nonumber\\
&\quad\qquad+\left(\frac{564\zeta_3}{5}-\frac{1700563}{54000}
-\frac{16291 \pi^2}{2025}\right)C_A C_F +
\left(-\frac{56\zeta_3}{9}+\frac{219077}{97200}+\frac{2411 \pi^2}{2025}\right)C_F^2\bigg]
\nonumber\\
&\quad+n_f^2 T_F^2 \bigg[\left(-\frac{256\zeta_3}{9}
-\frac{73076}{10125}+\frac{320 \pi^2}{81}\right)C_A 
+\left(-\frac{2611}{40500}-\frac{392 \pi^2}{405}\right)C_F\bigg] \ ,
\nonumber\\
\gamma_{gg}^{(2)}(4)&=
\left(-\frac{3752\zeta_3}{25}+\frac{1069405919}{1350000}-\frac{171289 \pi^2}{1125}
+\frac{28 \pi^4}{3}\right)C_A^3 
\nonumber\\
&\quad+n_f T_F \bigg[
\left(-\frac{59068\zeta_3}{225}+\frac{129284923}{1215000}-\frac{30316 \pi^2}{2025}
+\frac{80 \pi^4}{27}\right)C_A^2
\nonumber\\
&\quad\qquad+\left(\frac{5488\zeta_3}{45}
\!-\!\frac{188283293}{3037500}\!-\!\frac{2158 \pi^2}{375}\right)C_A C_F\!+\!
\left(\frac{27742123}{12150000}\!-\!\frac{704\zeta_3}{225}
\!+\!\frac{4037 \pi^2}{10125}\right)C_F^2\bigg]
\nonumber\\
&\quad+n_f^2 T_F^2 \bigg[\left(-\frac{256\zeta_3}{9}
-\frac{71341}{6750}+\frac{320 \pi^2}{81}\right)C_A 
+\left(-\frac{165553}{67500}-\frac{968 \pi^2}{2025}\right)C_F\bigg] \ ,
\nonumber\\
&
\nonumber\\
\gamma_{qg}^{(2)}(2)&=
T_F \bigg\{\left(\frac{128\zeta_3}{9}-\frac{6232}{243}+\frac{1066 \pi^2}{81}
-\frac{40 \pi^4}{27}\right)C_A^2 
\nonumber\\
&\quad\qquad+\left(-\frac{56\zeta_3}{9}-\frac{2681}{243}
+\frac{380 \pi^2}{81}\right)C_A C_F 
+\left(\frac{32\zeta_3}{9}-\frac{10570}{243}+\frac{176 \pi^2}{81}\right)C_F^2 
\nonumber\\
&\quad\qquad+n_f T_F\bigg[\left(\frac{128\zeta_3}{9}+\frac{4}{27}
-\frac{160 \pi^2}{81}\right)C_A 
+\left(\frac{128 \pi^2}{81}-\frac{82}{9}\right)C_F\bigg] \bigg\} \ ,
\nonumber\\
\gamma_{qg}^{(2)}(3)&=
T_F \bigg\{\left(\frac{343\zeta_3}{45}-\frac{1795237}{1944000}
+\frac{333019 \pi^2}{81000}-\frac{14 \pi^4}{25}\right)C_A^2 
\nonumber\\
&\quad\qquad+\left(\frac{6208\zeta_3}{75}-\frac{3607891}{38880}
+\frac{24821 \pi^2}{8100}-\frac{7 \pi^4}{75}\right)C_A C_F 
\nonumber\\
&\quad\qquad+\left(-\frac{26102\zeta_3}{225}
+\frac{9397651}{97200}-\frac{1021 \pi^2}{675}+\frac{224 \pi^4}{675}\right)C_F^2 
\nonumber\\
&\quad\qquad+n_f T_F \bigg[\left(\frac{1215691}{60750}\!-\!\frac{56\zeta_3}{9}
\!-\!\frac{3616 \pi^2}{2025}\right)C_A\!+\!\left(\frac{3584 \pi^2}{2025}
\!-\!\frac{10657}{2025}\right)C_F\!-\!\frac{688}{1125}n_f T_F\bigg] \bigg\} ,
\nonumber\\
\gamma_{qg}^{(2)}(4)&=
T_F \bigg\{\left(\frac{1004\zeta_3}{225}-\frac{140682763}{6075000}
+\frac{94231 \pi^2}{10125}-\frac{22 \pi^4}{27}\right)C_A^2 
\nonumber\\
&\quad\qquad+\left(\frac{6503\zeta_3}{225}
-\frac{509985949}{24300000}+\frac{7003 \pi^2}{4500}\right)C_A C_F 
\nonumber\\
&\quad\qquad+
\left(\frac{622\zeta_3}{225}-\frac{2412861131}{48600000}
+\frac{79361 \pi^2}{40500}\right)C_F^2
\nonumber\\
&\quad\qquad+n_f T_F \bigg[\left(\frac{352\zeta_3}{45}\!-\!\frac{51449}{4500}
\!-\!\frac{116 \pi^2}{405}\right)C_A\!+\!\left(\frac{3454 \pi^2}{2025}
\!-\!\frac{915539}{150000}\right)C_F\!-\!\frac{344}{375}n_f T_F\bigg\} \ ,
\nonumber\\
&
\nonumber\\
\gamma_{gq}^{(2)}(2)&=
\left(-\frac{64\zeta_3}{3}-\frac{20920}{243}\right)C_A^2 C_F 
+\left(-\frac{2464\zeta_3}{9}-\frac{6608}{243}+\frac{1216 \pi^2}{27}
+\frac{32 \pi^4}{27}\right)C_A C_F^2
\nonumber\\
&\quad+n_f T_F \bigg[\left(\frac{1024\zeta_3}{9}
-\frac{110}{81}-\frac{296 \pi^2}{81}\right)C_A C_F 
+\left(-\frac{128\zeta_3}{9}-\frac{4562}{81}+\frac{32 \pi^2}{27}\right)C_F^2\bigg]
\nonumber\\
&\quad+\left(320\zeta_3-\frac{54556}{243}+\frac{3632 \pi^2}{81}
-\frac{64 \pi^4}{9}\right)C_F^3 \ ,
\nonumber\\
\gamma_{gq}^{(2)}(3)&=
\left(-\frac{2791\zeta_3}{90}-\frac{17093053}{777600}-\frac{50593\pi^2}{3600}
+\frac{98 \pi^4}{135}\right)C_A^2 C_F 
\nonumber\\
&\quad+n_f T_F \bigg[
\left(\frac{364\zeta_3}{9}+\frac{246767}{30375}-\frac{73 \pi^2}{81}\right)C_A C_F
+\left(-\frac{56\zeta_3}{9}-\frac{419593}{40500}+\frac{4 \pi^2}{27}\right)C_F^2\bigg]
\nonumber\\
&\quad+\left(-\frac{3029\zeta_3}{9}+\frac{63294389}{388800}
+\frac{123773 \pi^2}{5400}+\frac{511 \pi^4}{270}\right)C_A C_F^2 
\nonumber\\
&\quad+\left(\frac{2533\zeta_3}{9}-\frac{647639}{3888}+\frac{3193\pi^2}{324}
-\frac{154 \pi^4}{45}\right)C_F^3 \ ,
\nonumber\\
\gamma_{gq}^{(2)}(4)&=
\left(\frac{6448\zeta_3}{75}-\frac{2010250477}{12150000}
-\frac{5449\pi^2}{1125}\right)C_A^2 C_F 
\nonumber\\
&\quad+n_f T_F \bigg[\left(\frac{1408\zeta_3}{45}+\frac{2334509}{202500}
-\frac{3736 \pi^2}{2025}\right)C_A C_F+
\left(\frac{152\pi^2}{675}-\frac{176\zeta_3}{45}-\frac{14837573}{2025000}\right)C_F^2\bigg]
\nonumber\\
&\quad+\left(-\frac{31346\zeta_3}{225}-\frac{1694499413}{24300000}
+\frac{234407 \pi^2}{6750}+\frac{44 \pi^4}{135}\right)C_A C_F^2 
\nonumber\\
&\quad+\left(\frac{1796\zeta_3}{15}-\frac{1061823161}{24300000}
+\frac{39634 \pi^2}{10125}-\frac{88 \pi^4}{45}\right)C_F^3 \ ,
\nonumber\\
&
\nonumber\\
\gamma_{qq}^{(2)}(2)&=
\left(\frac{3079\zeta_3}{9}+4 \pi^2\zeta_3+56\zeta_5+\frac{58853}{1944}
-\frac{485 \pi^2}{36}-\frac{857 \pi^4}{270}\right)C_A^2 C_F 
\nonumber\\
&\quad+n_f T_F\bigg[\left(\frac{34\pi^4}{135}-\frac{1088\zeta_3}{9}
-\frac{3616}{243}+6\pi^2\right)C_A C_F 
\nonumber\\
&\quad\qquad
+\left(\frac{1280\zeta_3}{9}-\frac{20680}{243}+\frac{20 \pi^2}{9}
-\frac{68 \pi^4}{135}\right)C_F^2
+\frac{448}{243}C_F T_F-\frac{896}{243} C_F n_f T_F\bigg]
\nonumber\\
&\quad+T_F
\bigg[\left(-\frac{320\zeta_3}{9}+\frac{3293}{243}+\frac{148 \pi^2}{81}\right)C_A C_F +\left(\frac{128\zeta_3}{9}+\frac{14543}{243}-\frac{208 \pi^2}{27}\right)C_F^2\bigg]
\nonumber\\
&\quad+\left(-\frac{5708\zeta_3}{9}-\frac{52 \pi^2\zeta_3}{3}
-216\zeta_5+\frac{163075}{486}
-\frac{3938 \pi^2}{81}+\frac{532 \pi^4}{45}\right)C_A C_F^2 
\nonumber\\
&\quad+\left(\frac{1916\zeta_3}{9}+\frac{56 \pi^2\zeta_3}{3}
+208\zeta_5-\frac{82099}{486}+\frac{1313 \pi^2}{81}-\frac{838 \pi^4}{135}\right)C_F^3 
\nonumber\\
&\quad+\left(\frac{4\zeta_3}{9}
-\frac{758}{243}+\frac{28\pi^2}{81}\right)\frac{d_{abc}d_{abc}}{N_c} \ ,
\nonumber\\
\gamma_{qq}^{(2)}(3)&=
\left(\frac{16483\zeta_3}{36}+4 \pi^2\zeta_3+56\zeta_5+\frac{508201}{62208}
-\frac{13105 \pi^2}{864}-\frac{2083 \pi^4}{540}\right)C_A^2 C_F 
\nonumber\\
&\quad+n_f T_F \bigg[\left(-\frac{1448\zeta_3}{9}-\frac{45515}{1944}
+\frac{437 \pi^2}{54}+\frac{34 \pi^4}{135}\right)C_A C_F 
\nonumber\\
&\quad\qquad+\left(\frac{1496\zeta_3}{9}-\frac{568813}{3888}+\frac{173\pi^2}{27}
-\frac{68 \pi^4}{135}\right)C_F^2 
+\frac{324853}{243000}C_F T_F-\frac{2569}{486} C_F n_f T_F\bigg]
\nonumber\\
&\quad+T_F \bigg[\left(-\frac{7\zeta_3}{5}-\frac{10843531}{1944000}
+\frac{15871 \pi^2}{16200}\right)C_A C_F +
\left(\frac{196\zeta_3}{45}+\frac{1796579}{97200}
-\frac{3167 \pi^2}{1350}\right)C_F^2\bigg]
\nonumber\\
&\quad+\left(-\frac{7247\zeta_3}{9}-\frac{52\pi^2\zeta_3}{3}-216\zeta_5
+\frac{1286017}{1944}-\frac{27689 \pi^2}{324}+\frac{646 \pi^4}{45}\right)C_A C_F^2 
\nonumber\\
&\quad+\left(\frac{2411\zeta_3}{9}+\frac{56\pi^2\zeta_3}{3}+208\zeta_5
-\frac{1997845}{7776}+\frac{8551 \pi^2}{648}-\frac{793\pi^4}{135}\right)C_F^3 
\nonumber\\
&\quad+\frac{205}{576}\frac{d_{abc}d_{abc}}{N_c} \ ,
\nonumber\\
\gamma_{qq}^{(2)}(4)&=
\left(\frac{19939\zeta_3}{36}+4 \pi^2\zeta_3+56\zeta_5-\frac{1265893697}{64800000}
-\frac{5470151 \pi^2}{324000}-\frac{1303 \pi^4}{300}\right)C_A^2 C_F 
\nonumber\\
&\quad+n_f T_F \bigg[\left(-\frac{568\zeta_3}{3}-\frac{34512043}{1215000}
+\frac{4319 \pi^2}{450}+\frac{34 \pi^4}{135}\right)C_A C_F
\nonumber\\
&\quad\qquad+\left(\frac{8248\zeta_3}{45}\!-\!\frac{94188089}{486000}
\!+\!\frac{2119\pi^2}{225}\!-\!\frac{68 \pi^4}{135}\right)C_F^2 
+\frac{236357}{243000}C_F T_F-\frac{384277}{60750}C_F n_f T_F\bigg]
\nonumber\\
&\quad+T_F\bigg[\left(\frac{2893382}{759375}\!-\!\frac{242\zeta_3}{45}
\!+\!\frac{4246\pi^2}{10125}\right)C_A C_F 
+\left(\frac{484\zeta_3}{225}+\frac{125062003}{12150000}
-\frac{8407 \pi^2}{6750}\right)C_F^2\bigg]
\nonumber\\
&\quad+\left(-\frac{43099\zeta_3}{45}-\frac{52\pi^2\zeta_3}{3}-216\zeta_5
+\frac{1856972509}{2025000}-\frac{4455181 \pi^2}{40500}
+\frac{10906 \pi^4}{675}\right)C_A C_F^2
\nonumber\\
&\quad+\left(\frac{2827\zeta_3}{9}+\frac{56 \pi^2\zeta_3}{3}+208\zeta_5
-\frac{7289888977}{24300000}+\frac{700843 \pi^2}{81000}-\frac{761 \pi^4}{135}\right)C_F^3
\nonumber\\
&\quad+
\left(\frac{11\zeta_3}{100}-\frac{183166273}{194400000}+\frac{32767\pi^2}{324000}\right)
\frac{d_{abc}d_{abc}}{N_c} \ ,
\nonumber\\
&
\nonumber\\
\gamma_{\bar{q}q}^{(2)}(2)&=
\left(-\frac{2887\zeta_3}{9}-4 \pi^2\zeta_3-56\zeta_5+\frac{36169}{648}
+\frac{485 \pi^2}{36}+\frac{857 \pi^4}{270}\right)C_A^2 C_F
\nn \\
&\quad+n_f T_F\bigg[\left(\frac{704\zeta_3}{9}-\frac{880}{81}-6\pi^2
-\frac{34 \pi^4}{135}\right)C_A C_F 
\nonumber\\
&\quad\qquad
+\left(\frac{1760}{81}-\frac{1408\zeta_3}{9}+12 \pi^2+\frac{68 \pi^4}{135}\right)C_F^2 
+\frac{448}{243}C_F T_F\bigg]
\nonumber\\
&\quad+T_F \bigg[\left(-\frac{320\zeta_3}{9}+\frac{3293}{243}
+\frac{148 \pi^2}{81}\right)C_A C_F +\left(\frac{128\zeta_3}{9}+\frac{14543}{243}
-\frac{208 \pi^2}{27}\right)C_F^2\bigg]
\nonumber\\
&\quad+\left(908\zeta_3+\frac{52 \pi^2\zeta_3}{3}+216\zeta_5
-\frac{16651}{54}+\frac{290 \pi^2}{81}-\frac{1756\pi^4}{135}\right)C_A C_F^2
\nonumber\\
&\quad+\left(-\frac{4796\zeta_3}{9}-\frac{56 \pi^2\zeta_3}{3}-208\zeta_5
+\frac{63737}{162}-\frac{4945 \pi^2}{81}+\frac{1798 \pi^4}{135}\right)C_F^3
\nonumber\\
&\quad+\left(-\frac{4\zeta_3}{9}+\frac{758}{243}-\frac{28\pi^2}{81}\right)
\frac{d_{abc}d_{abc}}{N_c} \ ,
\nonumber\\
\gamma_{\bar{q}q}^{(2)}(3)&=
\left(\frac{15823\zeta_3}{36}+4 \pi^2\zeta_3+56\zeta_5-\frac{2202421}{20736}
-\frac{13105 \pi^2}{864}-\frac{2083 \pi^4}{540}\right)C_A^2 C_F 
\nonumber\\
&\quad+n_f T_F \bigg[\left(-\frac{848\zeta_3}{9}+\frac{2789}{324}
+\frac{437 \pi^2}{54}+\frac{34 \pi^4}{135}\right)C_A C_F 
\nonumber\\
&\quad\qquad
+\left(\frac{1696\zeta_3}{9}-\frac{2789}{162}-\frac{437\pi^2}{27}
-\frac{68\pi^4}{135}\right)C_F^2 
+\frac{324853}{243000} C_F T_F\bigg]
\nonumber\\
&\quad+T_F \bigg[\left(\frac{15871\pi^2}{16200}-\frac{7\zeta_3}{5}
-\frac{10843531}{1944000}\right)C_A C_F +
\left(\frac{196\zeta_3}{45}+\frac{1796579}{97200}-\frac{3167 \pi^2}{1350}\right)C_F^2
\bigg]
\nonumber\\
&\quad+\left(-1353\zeta_3-\frac{52 \pi^2\zeta_3}{3}-216\zeta_5
+\frac{3347233}{5184}-\frac{5729 \pi^2}{324}+\frac{2188 \pi^4}{135}\right)C_A C_F^2 
\nonumber\\
&\quad+\left(\frac{8531\zeta_3}{9}+\frac{56 \pi^2\zeta_3}{3}+208\zeta_5
-\frac{4492045}{5184}+\frac{62231 \pi^2}{648}-\frac{2293 \pi^4}{135}\right)C_F^3 
-\frac{205}{576} \frac{d_{abc}d_{abc}}{N_c} \ ,
\nonumber\\
\gamma_{\bar{q}q}^{(2)}(4)&=
\left(-\frac{481207\zeta_3}{900}-4 \pi^2\zeta_3-56\zeta_5
+\frac{31010955691}{194400000}+\frac{5470151 \pi^2}{324000}
+\frac{1303 \pi^4}{300}\right)C_A^2 C_F 
\nonumber\\
&\quad+n_f T_F \bigg[\left(\frac{528\zeta_3}{5}-\frac{1583477}{202500}
-\frac{4319 \pi^2}{450}-\frac{34 \pi^4}{135}\right)C_A C_F 
\nonumber\\
&\quad\qquad+\left(-\frac{1056\zeta_3}{5}+\frac{1583477}{101250}
+\frac{4319 \pi^2}{225}+\frac{68 \pi^4}{135}\right)C_F^2 
+\frac{236357}{243000}C_F T_F\bigg]
\nonumber\\
&\quad+T_F\bigg[\left(\frac{2893382}{759375}-\frac{242\zeta_3}{45}
+\frac{4246 \pi^2}{10125}\right)C_A C_F +
\left(\frac{484\zeta_3}{225}+\frac{125062003}{12150000}
-\frac{8407 \pi^2}{6750}\right)C_F^2\bigg]
\nonumber\\
&\quad+\left(\frac{387961\zeta_3}{225}+\frac{52\pi^2\zeta_3}{3}+216\zeta_5
-\frac{46783579631}{48600000}+\frac{1058821 \pi^2}{40500}
-\frac{12476 \pi^4}{675}\right)C_A C_F^2 
\nonumber\\
&\quad+\left(-\frac{58943\zeta_3}{45}-\frac{56 \pi^2\zeta_3}{3}-208\zeta_5
+\frac{20852067857}{16200000}-\frac{647029 \pi^2}{5400}+\frac{529 \pi^4}{27}\right)C_F^3
\nonumber\\
&\quad+\left(-\frac{11\zeta_3}{100}+\frac{183166273}{194400000}
-\frac{32767 \pi^2}{324000}\right) \frac{d_{abc}d_{abc}}{N_c} \ ,
\nonumber\\
&
\nonumber\\
\gamma_{Qq}^{(2)}(2)&=
C_F T_F \bigg[\left(-\frac{320\zeta_3}{9}+\frac{3293}{243}+\frac{148\pi^2}{81}\right)C_A
\nonumber\\
&\quad\qquad
+\left(\frac{128\zeta_3}{9}+\frac{14543}{243}-\frac{208 \pi^2}{27}\right)C_F 
+\frac{448}{243} n_f T_F\bigg]
\nonumber\\
&\quad
+\left(\frac{4\zeta_3}{9}-\frac{758}{243}+\frac{28 \pi^2}{81}\right)
\frac{d_{abc}d_{abc}}{N_c} \ ,
\nonumber\\
\gamma_{Qq}^{(2)}(3)&=
C_F T_F \bigg[\left(-\frac{7\zeta_3}{5}-\frac{10843531}{1944000}
+\frac{15871 \pi^2}{16200}\right)C_A
\nonumber\\
&\quad\qquad
+\left(\frac{196\zeta_3}{45}+\frac{1796579}{97200}-\frac{3167 \pi^2}{1350}\right)C_F
+\frac{324853}{243000}n_f T_F\bigg]
\nonumber\\
&\quad
+\frac{205}{576}\frac{d_{abc}d_{abc}}{N_c} \ ,
\nonumber\\
\gamma_{Qq}^{(2)}(4)&=
C_F T_F \bigg[\left(-\frac{242\zeta_3}{45}+\frac{2893382}{759375}
+\frac{4246 \pi^2}{10125}\right)C_A 
\nonumber\\
&\quad\qquad
+\left(\frac{484\zeta_3}{225}+\frac{125062003}{12150000}-\frac{8407\pi^2}{6750}\right)C_F
+\frac{236357}{243000}n_f T_F\bigg]
\nonumber\\
&\quad
+\left(\frac{11\zeta_3}{100}-\frac{183166273}{194400000}+\frac{32767 \pi^2}{324000}\right) 
\frac{d_{abc}d_{abc}}{N_c} \ ,
\end{align}
where
\begin{align}
d_{abc}d_{abc}
&=2C_F(C_A^2-4)
=\frac{(N^2-4)(N^2-1)}{N}
\,.\end{align}
Along with the DGLAP anomalous dimensions, we also require the $\beta$ function, which we expand in powers of $\alpha_s$ as
\begin{align} \label{eq:betafunction}
\beta(\alpha_s) =
- 2 \alpha_s \sum_{n=0}^\infty \beta_n\Bigl(\frac{\alpha_s}{4\pi}\Bigr)^{n+1}
\ .
\end{align}
Up to three-loop order in the $\overline {\rm MS}$ scheme, the coefficients of the $\beta$ function are~\cite{Tarasov:1980au, Larin:1993tp}
\begin{align} \label{eq:cusp}
\beta_0 &= \frac{11}{3}\,C_A -\frac{4}{3}\,T_F\,n_f
\,, \qquad
\beta_1 = \frac{34}{3}\,C_A^2  - \Bigl(\frac{20}{3}\,C_A\, + 4 C_F\Bigr)\, T_F\,n_f
\,, \nn\\
\beta_2 &=
\frac{2857}{54}\,C_A^3 + \Bigl(C_F^2 - \frac{205}{18}\,C_F C_A
 - \frac{1415}{54}\,C_A^2 \Bigr)\, 2T_F\,n_f
 + \Bigl(\frac{11}{9}\, C_F + \frac{79}{54}\, C_A \Bigr)\, 4T_F^2\,n_f^2\,.
\end{align}

\bibliographystyle{JHEP}
\bibliography{spinning_gluon.bib}

\end{document}